\documentclass[aps,prl,epsfigure,twocolumn,superscriptaddress]{revtex4-2}
\usepackage[colorlinks=true,linkcolor=blue,urlcolor=blue,citecolor=blue,pdfusetitle]{hyperref}
\usepackage[utf8]{inputenc}
\usepackage[english]{babel}
\usepackage{amsmath}
\usepackage[caption = false]{subfig}
\usepackage{graphicx,epstopdf}
\usepackage{threeparttable} 
\usepackage{blindtext}
\usepackage[table,xcdraw]{xcolor}
\usepackage{lipsum}
\usepackage{amsfonts}
\usepackage{bbm}
\usepackage{ulem}
\usepackage{amssymb}
\usepackage{enumerate}
\usepackage{color}
\usepackage{latexsym}
\usepackage{physics}
\usepackage{times,txfonts}

\newcommand{\Ecal}{\mathcal{E}}

\newcommand{\Lcal}{\mathcal{L}}

\newcommand{\Pcal}{\mathcal{P}}

\newcommand{\1}{\mathbbm{1}}

\usepackage[bb=boondox]{mathalfa}

\newcommand{\SubFig}[2]{\ref{#1}{\color{blue}#2}}

\definecolor{bluePoli}{cmyk}{0.4,0.1,0,0.4}
\definecolor{blueGreen}{RGB}{44, 120, 247}
\definecolor{brickred}{rgb}{0.8, 0.25, 0.33}
\definecolor{darkred}{RGB}{204, 0, 0}
\definecolor{darkgreen}{RGB}{0, 102, 50}
\definecolor{darkblue}{RGB}{0, 76, 153}
\definecolor{mygold}{RGB}{255, 128, 32}
\definecolor{mypurple}{RGB}{178, 105, 252}
\definecolor{myorange}{RGB}{204, 102, 0}

\newcommand{\RefA}[1]{{\color{black}#1}}
\newcommand{\RefB}[1]{{\color{black}#1}}
\newcommand{\RefC}[1]{{\color{black}#1}}
\newcommand{\RefCnew}[1]{{\color{black}#1}}

\newcommand{\siqse}{Shenzhen Institute for Quantum Science and Engineering, Southern University of Science and Technology, Shenzhen, Guangdong 518055, China}

\newcommand{\gdpkl}{Guangdong Provincial Key Laboratory of Quantum Science and Engineering, Southern University of Science and Technology, Shenzhen, Guangdong 518055, China}
\newcommand{\szkl}{International Quantum Academy, Futian District, Shenzhen, Guangdong 518048, China}

\newcommand{\CSIC}{Instituto de Física Fundamental, Consejo Superior de Investigaciones Científicas, Calle Serrano 113b, 28006 Madrid, Spain}

\newcommand{\UFF}{Instituto de F\'{i}sica, Universidade Federal Fluminense, Av. Gal. Milton Tavares de Souza s/n, Gragoat\'{a}, 24210-346 Niter\'{o}i, Rio de Janeiro, Brazil}
\newcommand{\HFNL}{Shenzhen Branch, Hefei National Laboratory, Shenzhen 518048, China}

\newcommand{\Title}{Quantum Charging Advantage in Superconducting Solid-State Batteries}

\usepackage{etoolbox}

\usepackage{orcidlink}
\begin{document}
	
	\title{\Title}

	\author{Chang-Kang Hu}
	\thanks{These authors contributed equally to this work.}
	\affiliation{\szkl}
	
	\author{Chilong Liu}
	\thanks{These authors contributed equally to this work.}
	\affiliation{\szkl}\affiliation{\gdpkl}
	
	\author{Jingchao Zhao}
	\affiliation{\szkl}\affiliation{\gdpkl}
	
	\author{Liuzhu Zhong}
	\affiliation{\szkl}\affiliation{\gdpkl}
	
	\author{Yuxuan Zhou}
	\affiliation{\szkl}
	
	\author{Mingze Liu}
	\affiliation{\szkl}\affiliation{\gdpkl}
	
	\author{Haolan Yuan}
	\affiliation{\szkl}\affiliation{\gdpkl}
	
	\author{Yongchang Lin}
	\affiliation{\szkl}\affiliation{\gdpkl}
	
	\author{Yue Xu}
	\affiliation{\szkl}\affiliation{\gdpkl}
	
	\author{Guantian Hu}
	\affiliation{\szkl}\affiliation{\gdpkl}
	
	\author{Guixu Xie}
	\affiliation{\szkl}\affiliation{\gdpkl}
	
	\author{Zixing Liu}
	\affiliation{\szkl}
	
	\author{Ruiyang Zhou}
	\affiliation{\szkl}\affiliation{\gdpkl}
	
	\author{Yougui Ri}
	\affiliation{\szkl}\affiliation{\gdpkl}
	
	\author{Wenxuan Zhang}
	\affiliation{\szkl}\affiliation{\gdpkl}
	
	\author{Ruicheng Deng}
	\affiliation{\szkl}\affiliation{\gdpkl}
	
	\author{Andreia Saguia~\orcidlink{0000-0003-0403-4358}}
	\affiliation{\UFF}
	\author{Xiayu Linpeng}
	\affiliation{\szkl}\affiliation{\gdpkl}
	
	\author{Marcelo S. Sarandy~\orcidlink{0000-0003-0910-4407}}
	\affiliation{\UFF}
	
	\author{Song Liu}
	\affiliation{\szkl}\affiliation{\HFNL}
	
	\author{Alan C. Santos~\orcidlink{0000-0002-6989-7958}}
	\email{ac\_santos@iff.csic.es}
	\affiliation{\CSIC}
	
	\author{Dian Tan}
	\email{tandian@iqasz.cn}
	\affiliation{\szkl}\affiliation{\HFNL}
	
	\author{Dapeng Yu}
	\affiliation{\szkl}\affiliation{\HFNL}

	\begin{abstract}
		Quantum battery, as a novel energy storage device, offers the potential for unprecedented efficiency and performance beyond the capabilities of classical systems, with broad implications for future quantum technologies. Here, we experimentally \RefC{demonstrate quantum charging advantage (QCA)} in a scalable solid-state quantum battery. More specifically, we show how double-excitation Hamiltonians for two-level systems promote scalable QCA \RefB{with standard methods.} We effectively implement the collective evolution of quantum systems with 2 up to 12 battery cells in a superconducting quantum processor, and study the performance of quantum charging compared to its uncorrelated classical counterpart. The model considered is a linear chain of superconducting transmon qubits with only \textit{nearest-neighbor} and \textit{pairwise} interactions, which constitute the simplest model of a multi-cell quantum battery. Our results empirically realize substantial QCA without the necessity of adopting long-range and many-body interactions \RefB{ and showcase the quantum features of the QB charging processes with measurements of non-zero coherent ergotropy, incoherent ergotropy and entanglement,} revealing a promising prospect for further developments of efficient and experimentally feasible protocols for QCA.
	\end{abstract}

	\maketitle
	
	\emph{Introduction.--} The current stage of quantum technologies requires a quantum energy initiative~\cite{Alexia:22,campbell2025}. In this regard, energy storage quantum devices, also known as quantum batteries (QBs), have been extensively investigated over the past recent years~\cite{Alicki:13,Crescente:20,Quan:23,Campaioli:24}. Theoretical proposals have been developed to predict their performance in platforms of spin systems~\cite{Le:18,Grazi:24}, atomic emitters in cavity~\cite{Ferraro:18}, trapped-ion quantum simulators~\cite{Lai:24,zhang2024single}, among others (e.g., see Ref.~\cite{Campaioli:24} for a review on QBs). On the experimental side, charging, energy storage, and work extraction have been realized in a variety of architectures, such as spin systems~\cite{Joshi:22}, quantum dots~\cite{Maillette:23}, organic chemical systems~\cite{Quach:22,Cruz:22}, optical systems~\cite{Peng2023} and superconducting devices~\cite{Hu:22a,Li:25}. The intense interest in applications of QBs has led to a pivotal question: ``\textit{what is, in fact, quantum in the charging advantage of QBs?}". Indeed, this is  relevant to ascertain that a quantum device outperforms its classical counterpart, only exploiting quantum features, such as collective effects~\cite{Rossini:20}, entanglement,  and coherence~\cite{Shi:22,Alexia:20,Niu:24}. To address this point, remarkable progress have been done in order to define parameters and conditions able to \RefC{witness quantum charging advantage (QCA)}~\cite{PRL2017Binder,Kim:22, Lin2022}.
	
	Conditions to QCA take into account the charging performance and the energy cost related to the charging fields, constituting a robust framework to test the hypothetical advantage of QBs, with respect to their classical counterpart. \RefB{Of particular interest, the conditions established in Refs.~\cite{PRL2017Binder,Kim:22} takes into account the performance of quantum charging under an energetically fair scenario, where the quantum charging is allowed to use external fields with energy cost at most equal to the classical counterpart.} While the scaling \RefC{of QCA} in QBs has been extensively investigated at the theoretical level, no experimental evidence has substantiated these predictions yet. This gap largely stems from the challenge of engineering the complex many-body interactions required by such models to serve as the charging field. In fact, QCA has been predicted with many-body~\cite{Rossini:20,Kim:22}, non-linear interaction in bosonic systems~\cite{Andolina:24}, and long-ranged all-to-all interactions~\cite{Le:18}, with the optimal extensive power advantage with respect to the classical case   achieved in the limit of globally interacting systems~\cite{Binder:15,PRL2017Binder,Kim:22}. 
	
	\begin{figure}[t!]
		\includegraphics[width=\linewidth]{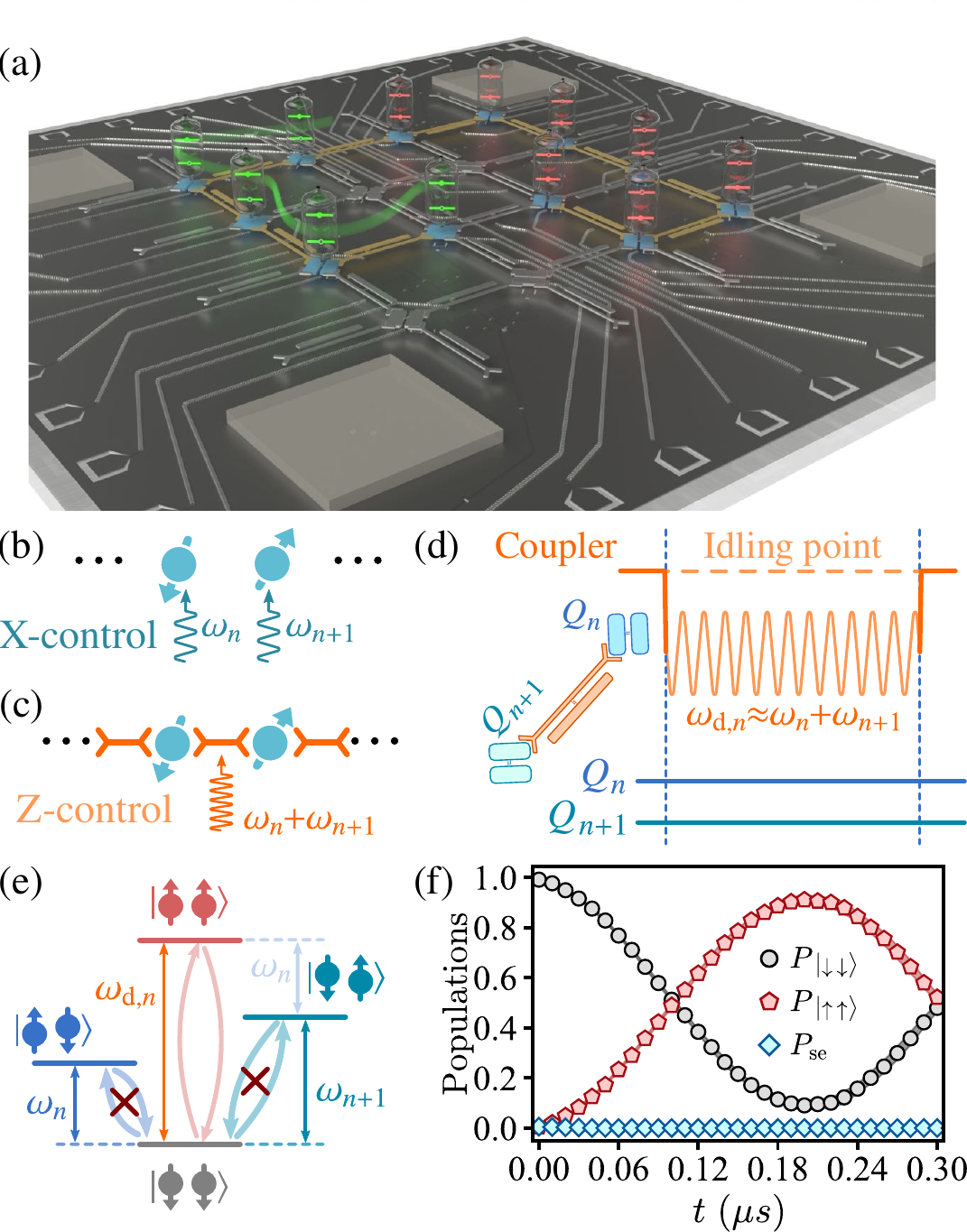}
		\caption{(a) Pictorial representation of our QB encoded in a 16-qubit lattice, with 12 cells activated in the system. The tunable qubit-qubit interactions control the number of cells during the quantum charging (green cells), while the rest of the system is not charged (red cells). (b) and (c) show the local drives used for classical and quantum charging processes, respectively. (d) Frequency profile for two neighboring qubits, and their corresponding coupler, to implement a two-photon transition by blocking transitions to single-excitation subspace, as depicted in Fig. (e). (f) Population dynamics for a two-cell QB showing low population in the single excitation subspace, $P_\mathrm{se} = P_\mathrm{\ket{\uparrow\downarrow}} + P_\mathrm{\ket{\downarrow\uparrow}}\ll 1$, while transitions $\ket{\downarrow\downarrow}\rightleftarrows\ket{\uparrow\uparrow}$ dominate the evolution, a clear witness of the anti-blockade-like mechanism through two-photon transition.}
		\label{Fig:Setup}
	\end{figure}
	
	In this work, we make quantum charging possible by introducing a charging protocol able to \RefC{reach QCA using} only nearest-neighbor pairwise interactions. \RefB{Here, we follow the standard methods for quantifying QCA by analyzing the energy cost and charging power enhancement~\cite{Campaioli:24,PRL2017Binder,Kim:22}}. We exploit the scalability of superconducting circuits as a viable platform for implementing multi-cell QBs, with each qubit acting as a battery cell, as shown in Fig.~\SubFig{Fig:Setup}{a}. In our device, we scale the number of battery cells from $N=2$ to $N=12$, so that the characterization of the quantum charging advantage is achieved through measurements of the maximum stored ergortopies, the average power over time, and the advantage parameter as a function of $N$. \RefB{Finally, we study the quantum features by measuring the coherent and incoherent ergotropies  and the entanglement that have raised during the charing process of our QBs.}
	
	\emph{Quantum charging protocol.--} The QB under consideration is composed of $N$ two-level cells, where each cell has ground $\ket{\downarrow}$ and excited $\ket{\uparrow}$ states energetically spaced by $\Ecal_{0} = \hbar \omega_{0}$, the maximum individual ergotropy capability~\cite{Yang:23}. This system can be seen as a chain of spin$-1/2$ particles with $N$ sites. The reference Hamiltonian reads $\hat{H}_{0} = \hbar \sum\nolimits_{n=1}^{N} \omega_{0}\hat{\sigma}^{+}_{n}\hat{\sigma}^{-}_{n}$, with $\hat{\sigma}^{+}_{n} = \ket{\uparrow}\bra{\downarrow} = \hat{\sigma}^{-\dagger}_{n}$. The QB can be charged either individually or collectively through the driving Hamiltonians $\hat{V}_{\mathrm{cl}} =  \sum_{n=1}^{N} \hbar\Omega \left(\hat{\sigma}^{+}_{n} + \hat{\sigma}^{-}_{n} \right)$ and
	\begin{align}
		\hat{V}_{\mathrm{qu}} =  \sum\nolimits_{n=1}^{N-1} \hbar g\left(\hat{\sigma}^{+}_{n}\hat{\sigma}^{+}_{n+1} + \hat{\sigma}^{-}_{n}\hat{\sigma}^{-}_{n+1}\right) , \label{Eq:HcollectiveFF} 
	\end{align}
	which correspond to the classical and quantum charging protocols, respectively. The parameter $\Omega$ is the classical driving strength and $g$ is the coupling strength between the cells $n$ and $n+1$.  \RefB{Note that the driving fields and Hamiltonians are different for the classical and quantum charging, yet they are compared under the same energetic constraints.} In the classical charging, $\hat{V}_{\mathrm{cl}}$, correlations and collective effects are not boost resources, as the dynamics is implemented through a resonant local drive applied to each battery cell, see Fig.~\SubFig{Fig:Setup}{b}. 
	
	Also, we introduce a theoretical quantum charging process by using Hamiltonian able to exploit collective effects through the nearest-neighbor double-excitation Hamiltonian $\hat{V}_{\mathrm{qu}}$~\cite{Ramanathan:03,Ramanathan:11}. Similarly to the classical charging field $\hat{V}_{\mathrm{cl}}$, the quantum charging $\hat{V}_{\mathrm{qu}}$ does not preserve the excitation number in the system and, consequently, it can be used to charge a QB starting from the fully empty energy state $\ket{\psi_{0}}$. By integrating the dynamics of the system, we show that the Hamiltonian $\hat{V}_{\mathrm{qu}}$ outperforms the classical charging, thus witnessing \RefC{of QCA}.  In the following, we focus on the experimental demonstration of this QCA, while the theoretical analysis is provided in the Supplementary Material~\cite{SM}.
	
	\emph{Experimental setup.--} 
	We begin by employing a 16-qubit lattice-structured superconducting quantum processor, as depicted in Fig.~\SubFig{Fig:Setup}{a}. All qubits are frequency tunable, allowing for a frequency tuning range of approximately $1.5$~GHz by applying DC signals through the flux lines. Pairwise interaction is engineered in our device through frequency-tunable couplers integrated between neighboring qubits, see Fig.~\SubFig{Fig:Setup}{c}. The pulse sequence for implementing the double-excitation Hamiltonian, Eq.~(\ref{Eq:HcollectiveFF}), 
	\RefB{
		with qubit-qubit bSWAP operations is detailed in Fig.~\SubFig{Fig:Setup}{d}. Through the local Z-line, we only manipulate the coupler between two adjacent qubits by applying a sinusoidal periodic modulated magnetic flux with a frequency $\omega_{\mathrm{d},n} \approx \omega_{n} + \omega_{n+1}$ close to the sum of the two neighboring qubit frequencies. Throughout the dynamics, the couplers are maintained in their ground states ~\cite{Marco:17,Ganzhorn:20,Bruno:22} and the effective Hamiltonian is strictly \RefC{an $N$-qubit} chain model.
	}
	We set the frequencies of the couplers near the interaction-off points of neighboring qubits, where the idling points are crucial for scalable charging. Moreover, the Hamiltonian $\hat{V}_{\mathrm{qu}}$ can be efficiently engineered if we induce a two-photon transition in a pair of cells, suppressing then the undesired transitions $\ket{\downarrow\downarrow} \leftrightarrows \{\ket{\uparrow\downarrow},\ket{\downarrow\uparrow}\}$ as illustrated in  Fig.~\SubFig{Fig:Setup}{e}. See~\cite{SM} for further details on the processor and the experimental setup.
	
	The experimentally obtained dynamics of a two-cell QB is shown in Fig.~\SubFig{Fig:Setup}{f}, highlighting the population evolution in the single- and two-excitation subspaces. The two-photon transition efficiently implements the required dynamics. Building on this, the scalability of the bSWAP operations on multi-cell QB is enabled by local flux control lines connected to each coupler, which allows for simultaneous application of periodic modulation to each coupler. By leveraging this capability, the quantum charging under the Hamiltonian $\hat{V}_{\mathrm{qu}}$ is \RefB{experimentally} implemented for systems with the number of cells ranging from 2 up to 12. \RefB{See~\cite{SM} for theoretical simulations on the scaling of the battery up to $N=21$}.

	\emph{Figure of merit for QCA.--} Given $\hat{\rho}(t)$ the state of the system and $\hat{H}_{0}$ the reference Hamiltonian, the ergotropy is defined as $\Ecal = \tr(\hat{\rho} \hat{H}_{0}) - \tr(\hat{\sigma}_{\rho} \hat{H}_{0})$, where $\hat{\sigma}_{\rho}$ represents the passive state of the system obtained by maximization over all unitary operations on $\hat{\rho}$~\cite{Allahverdyan:04}. In particular, for pure states $\hat{\rho}=\ket{\psi}\bra{\psi}$ and $H_0$ with vanishing ground-state energy, the ergotropy simplifies to $\Ecal = \bra{\psi}\hat{H}_{0}\ket{\psi}$. As shown in~\cite{SM}, our system is subject to decoherence and evolves governed by the Lindblad master equation. 
	From the ergotropy, we then obtain the average charging power using the standard definition $\bar{\Pcal} (\Delta t) = \Ecal (\Delta t) / \Delta t $, where $\Delta t$ is the charging time interval. The scalability of the charging process is characterized by the optimal charging power of the battery $\bar{\Pcal}^{\mathrm{opt}} (N) = \max_{\Delta t} \bar{\Pcal} (\Delta t)$, which depends on the number of qubits $N$ in the system.  To quantify the QCA, we \RefB{use the standard QCA parameter $\Gamma_{\mathrm{ad}}$ as \RefC{follows}~\cite{PRL2017Binder,Kim:22}}
	\begin{equation}
		\Gamma_{\mathrm{ad}} = \bar{\Pcal}^{\mathrm{opt}}_{\mathrm{qu}}  / \bar{\Pcal}^{\mathrm{opt}}_{\mathrm{cl}}  - 1 , \label{Eq:AdParameter}
	\end{equation}
	which measures the relative power enhancement of quantum charging over classical charging, with $\Gamma_{\mathrm{ad}} > 0$ indicating quantum charging advantage. \RefC{Although we focus on the average power as our primary metric in this discussion, analogous results are obtained when considering instantaneous power (see, e.g., the End Matter section).}
	
	\begin{figure}[t!]
		\includegraphics[width=\linewidth]{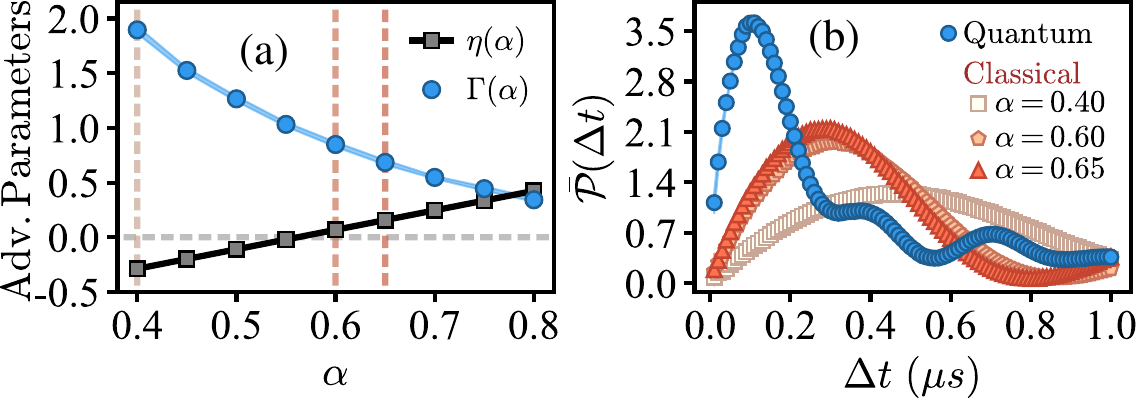}
		\caption{(a) Advantage parameters for a $5$-cell QB as a function of the ratio $\alpha = \Omega/g$, showing the transition from artificial energy powered \RefC{to QCA} (around $\alpha = 0.56$). (b) For the cases highlighted in (a), we show the classical average power over time, as a function of $\Delta t$, with the quantum charging power as reference. The experimentally measured coupling of the quantum charging is $g\approx 1.04 \times 2\pi$~MHz. \RefA{Except for the parameter $\eta (\alpha)$, all other datasets are based on experimental data.  Error bars indicate the statistical uncertainty, which is too small to display.}}
		\label{Fig:genuine_CQA_5qubits}
	\end{figure}
	
	The second key quantity for characterizing QCA is the \textit{driving potential} parameter~\cite{PRL2017Binder,Kim:22}. Such a parameter is defined for the classical and quantum charging as $v_{\mathrm{cl}}^{\mathrm{dv}} = ||\hat{V}_{\mathrm{cl}} - v^{\mathrm{min}}_{\mathrm{cl}}||$ and $v_{\mathrm{qu}}^{\mathrm{dv}} = ||\hat{V}_{\mathrm{qu}} - v^{\mathrm{min}}_{\mathrm{qu}}||$, where $v_{\mathrm{cl}}^{\mathrm{min}}$ and $v_{\mathrm{qu}}^{\mathrm{min}}$ are the minimum eigenvalue of the charging Hamiltonians $\hat{V}_{\mathrm{cl}}$ and 
	$\hat{V}_{\mathrm{qu}}$, respectively. As shown in Refs.~\cite{PRL2017Binder,Kim:22}, whenever $v_{\mathrm{cl}}^{\mathrm{dv}} \geq v_{\mathrm{qu}}^{\mathrm{dv}}$, the QB does not take advantage of extra resources during its charging process. Therefore, \RefC{identifying QCA} requires careful consideration of both power enhancement and energetic cost. To this end, we introduce the \textit{classical-quantum driving potential ratio} as
	\begin{equation}
		\eta = v^{\mathrm{dv}}_{\mathrm{cl}} / v^{\mathrm{dv}}_{\mathrm{qu}} - 1 ,
	\end{equation}
	\RefB{which accounts for differences in Hamiltonian norms by measuring the relative energetic cost of the quantum versus classical charging.}
	By introducing $\Gamma_{\mathrm{ad}}$ and $\eta$, we \RefB{follow a standard framework} for \RefC{identifying QCA}. We define that if $\Gamma_{\mathrm{ad}}(N)>0$ and $\eta(N) \geq 0$, \RefC{QCA} in the energy charging of the system is witnessed~\cite{PRL2017Binder,Kim:22}. This \RefB{well established condition} implies that the quantum charging power is enhanced with respect to the classical case, as one gets $\bar{\Pcal}^{\mathrm{opt}}_{\mathrm{qu}}>\bar{\Pcal}^{\mathrm{opt}}_{\mathrm{cl}}$, but without benefiting from additional energetic resources in the quantum driving Hamiltonian, since  $v_{\mathrm{cl}}^{\mathrm{dv}}\geq v_{\mathrm{qu}}^{\mathrm{dv}}$. 
	
	\emph{\RefC{QCA versus energy powered advantage}.--} To experimentally demonstrate the QCA behavior characterized by the parameters $\Gamma_{\mathrm{ad}}$ and $\eta$, we first implement the charging protocol for a QB composed of five cells. The effective dynamics of the quantum charging is given by the Hamiltonian in Eq.~\eqref{Eq:HcollectiveFF}, with the inter-cell coupling strength $g$ obtained from experimental data. We vary the strength $\Omega$ to exploit different regimes of classical charging. In fact, we can control how efficient and energetically demanding the classical charging is with respect to the quantum case through different choices of $\Omega$.
	Indeed, this is exhibited in  Fig.~\SubFig{Fig:genuine_CQA_5qubits}{a}, which shows the behavior of the QCA parameter $\eta$ and $\Gamma_{\mathrm{ad}}$, as a function of the ratio $\alpha = \Omega/g$. In the weak regime of the classical charging field strength (e.g. $\alpha \in \alpha_{I}  = [0.40,0.55]$), the parameter $\Gamma_{\mathrm{ad}}(\alpha_{I}) > 0$ indicates an enhanced quantum charging. However, since  $\eta(\alpha_{I}) < 0$ in this range, such enhancement is taking advantage of additional energy cost, and therefore it is an artificial energy powered QCA. On the other hand, by increasing $\Omega$ to higher values (e.g. $\alpha \in \alpha_{II}  = [0.60,0.80]$), the QCA parameter $\Gamma_{\mathrm{ad}}(\alpha_{II})$ is reduced, as $\Gamma_{\mathrm{ad}}(\alpha_{II}) < \Gamma_{\mathrm{ad}}(\alpha_{I})$, but $\Gamma_{\mathrm{ad}}(\alpha_{II})$ still indicates substantial quantum advantage in the interval $\alpha_{II}$. In this regime, one also gets $\eta(\alpha_{II}) > 0$, implying then
	\RefC{in a QCA}, with both conditions $\Gamma_{\mathrm{ad}}(\alpha_{II}) > 0$ and $\eta(\alpha_{II}) > 0$ satisfied. 
	
	\begin{figure}[t!]
		\includegraphics[width=\linewidth]{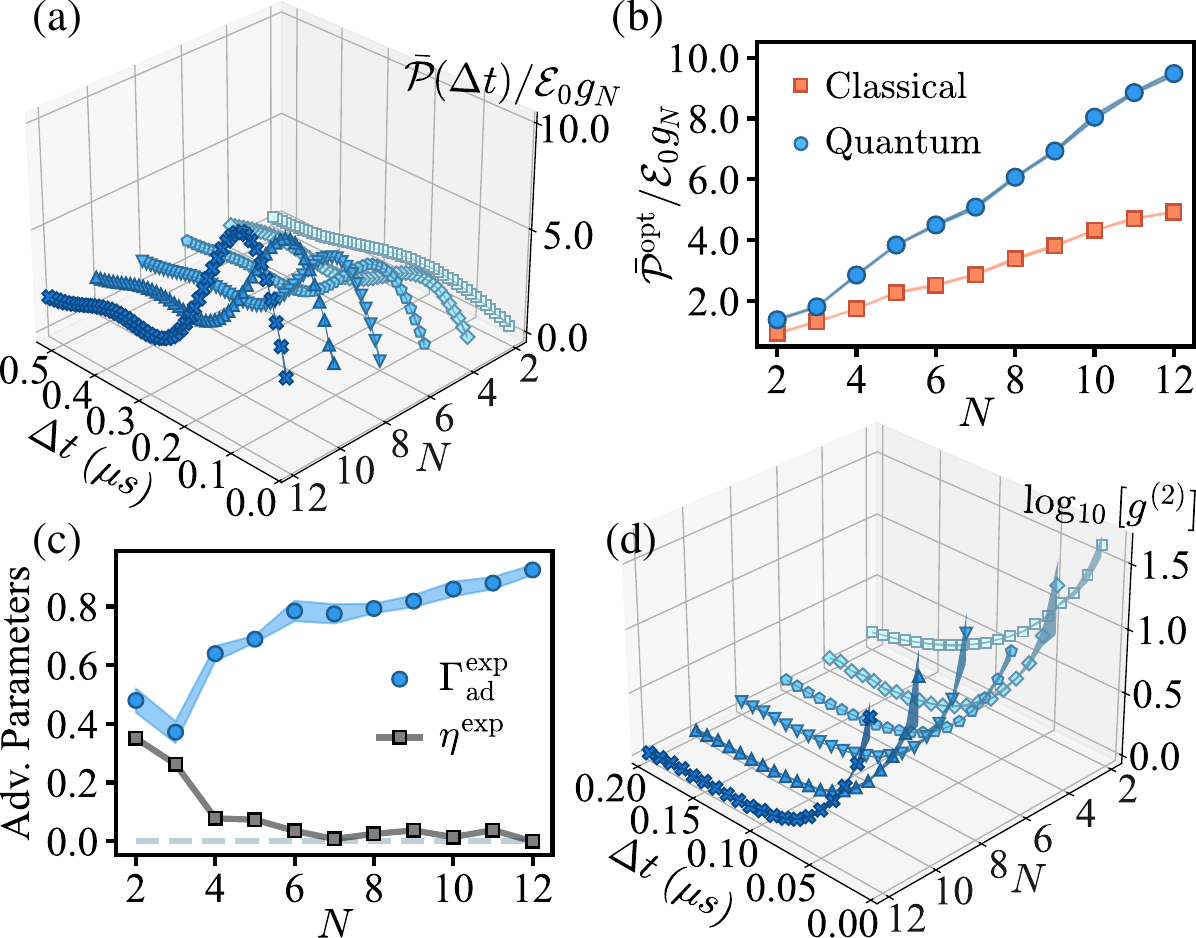}
		\caption{(a) Time-average quantum charging power for different values of $N$.  (b) The scaling of the maximum power for the quantum and classical protocols as a function the $N$. (c) Shows the behavior of the advantage parameters, used to \RefC{characterize QCA}. (d) Correlation function $g^{(2)}$ as a function of the charging time interval $\Delta t$ for different values of $N$, showing the giant-bunching in the optimal charging interval $\Delta t\leq 0.2~\mu$s. See~\cite{SM} for further details about the behavior of the parameters $\alpha$, $v_{\mathrm{qu}}$ and $v_{\mathrm{cl}}$ as function of $N$. \RefA{Error bars indicate the statistical uncertainty as shown in the shaded regions.}}
		\label{Fig:QCA_scaling}
	\end{figure}
	
	Furthermore, we show in Fig.~\SubFig{Fig:genuine_CQA_5qubits}{b} the comparison of the average power for quantum and classical charging, $\bar{\Pcal} (\Delta t)$, for the range choices discussed for the ratio $\alpha$ in Fig.~\SubFig{Fig:genuine_CQA_5qubits}{a}. The result clearly shows that the quantum charging can outperform classical charging, particularly in regimes where it incurs higher energetic costs. \RefB{We demonstrate that $||\hat{V}_{\mathrm{qu}}|| \leq ||\hat{V}_{\mathrm{cl}}||$, meaning that the observed advantage would in fact be even stronger compared to the case with exactly equal norms.} In light of this, we henceforth focus exclusively on parameter regimes where quantum charging satisfies \RefC{the QCA condition}. This can be efficiently ensured by choosing values of $\alpha$ such that $\eta(N) \geq 0$ for any number $N$ of cells in the QB.
	
	\emph{\RefC{Scaling of QCA.}--} We now implement the charging protocol for QBs with number of cells $N$ from 2 up to 12. To address the scaling \RefC{of QCA}, we begin by measuring the time-dependent performance of the protocol, quantified by the average charging power as a function of $\Delta t$ for different values of $N$, as displayed in Fig.~\SubFig{Fig:QCA_scaling}{a}. In the experiment, the effective coupling $g$ is not identical across different battery sizes~\cite{SM}, affecting then the coupling $g_{N}$ and the $\alpha_{N}$ ratio for a $N$-cell QB. To ensure a fair comparison under consistent energy cost, for each $N \in [2,12]$ we experimentally determine the quantities $\{g_{N},\alpha_{N}\}$, then obtaining the average power shown in Fig.~\SubFig{Fig:QCA_scaling}{a}~\footnote{It is worth mentioning that the average $\bar{g} = (1/12)\sum_{n=1}^{12} g_{n} \approx 1.03 \times 2\pi$~MHz, with standard deviation $\delta g \approx 31.52 \times 2\pi $~kHz.}. Notably, the maximum average power for $N\geq 4$ is consistently observed at $\Delta t_{\mathrm{max}} \approx 0.1~\mu\mathrm{s}$ suggesting an optimal time interval for charging~\cite{SM}.
	
	The optimal charging power with $N$ shows a better linear scaling behavior for the quantum charging compared to its classical counterpart, as shown in Fig.~\SubFig{Fig:QCA_scaling}{b}. We compare the classical and quantum charging protocols in Fig.~\SubFig{Fig:QCA_scaling}{c}, where the behavior of the advantage parameter $\Gamma_{\mathrm{ad}}(N)$ shows a clear and significant quantum charging enhancement as the battery size $N$ increases. In particular, this enhancement is compatible with the subextensive advantage possible for pairwise interactions, with $\bar{\Pcal}^{\mathrm{opt}}_{\mathrm{qu}}$ respecting the quadratic upper limit as a function of $N$~\cite{Kim:22}. Furthermore, by taking into account the driving potential ratio $\eta(N)$, we confirm that the condition $\eta(N) \geq 0$ is satisfied for any $N \in [2,12]$, thus establishing our charging protocol as \RefC{a realization of QCA}. 
	
	To understand the mechanism behind the advantage of our quantum charging protocol, we now show that it is closely related to an anti-blockade-like effect. To this end, we measured the atomic pairwise second-order correlation function~\cite{labuhn2016tunable,Bernien:17,Cidrim:20}
	\begin{align}
		g^{(2)}(t) = \frac{1}{N-1} \sum\nolimits_{j = 1}^{N-1} \frac{\langle \hat{n}_{j} \hat{n}_{j+1}\rangle}{\langle \hat{n}_{j}\rangle\langle \hat{n}_{j+1}\rangle}, \label{Eq:g2}
	\end{align}
	where $\langle \hat{X}\rangle = \tr(\rho(t)\hat{X})$ \RefC{and $\hat{n}_{j} = \hat{\sigma}_{j}^{+}\hat{\sigma}_{j}^{-}$}. The $g^{(2)}$-function serves as a witness to excitation blockade processes. Specifically, $g^{(2)}\ll 1$ indicates nearest-neighbor blockade, as observed in Rydberg atoms~\cite{labuhn2016tunable,Bernien:17} and low-lying excited neutral atoms~\cite{Cidrim:20}. In contrast, $g^{(2)} > 1$, is related to excitation-facilitating processes promoted by cooperative effects of the interacting system. For uncorrelated two-photon transitions, such as those expected in classical charging protocols, we obtain $g^{(2)} = 1$ at all times. This is because independent classical charging involves uncorrelated excitation processes, satisfying $\langle \hat{n}_{j} \hat{n}_{j+1}\rangle = \langle \hat{n}_{j}\rangle\langle \hat{n}_{j+1}\rangle$ for all $t > 0$, and hence yields $g_{\mathrm{cl}}^{(2)}(t) = 1$. The experimental results are shown in Fig.~\SubFig{Fig:QCA_scaling}{d} for QB with different number of cells $N$.
	The data show clear evidence of strong excitation facilitating dynamics during ergotropy injection within the time interval $\Delta t \leq 0.1~\mu\mathrm{s}$, which coincides with the required amount of time to reach the maximum average power. This behavior is even more accentuated for the case $N=2$, where the two-qubit system effectively forms a diatomic two-level configuration, similar to the anti-blockade effects observed in two-Rydberg atoms~\cite{Ates:07}. 
	
	\begin{figure}[t!]
		\includegraphics[width=\linewidth]{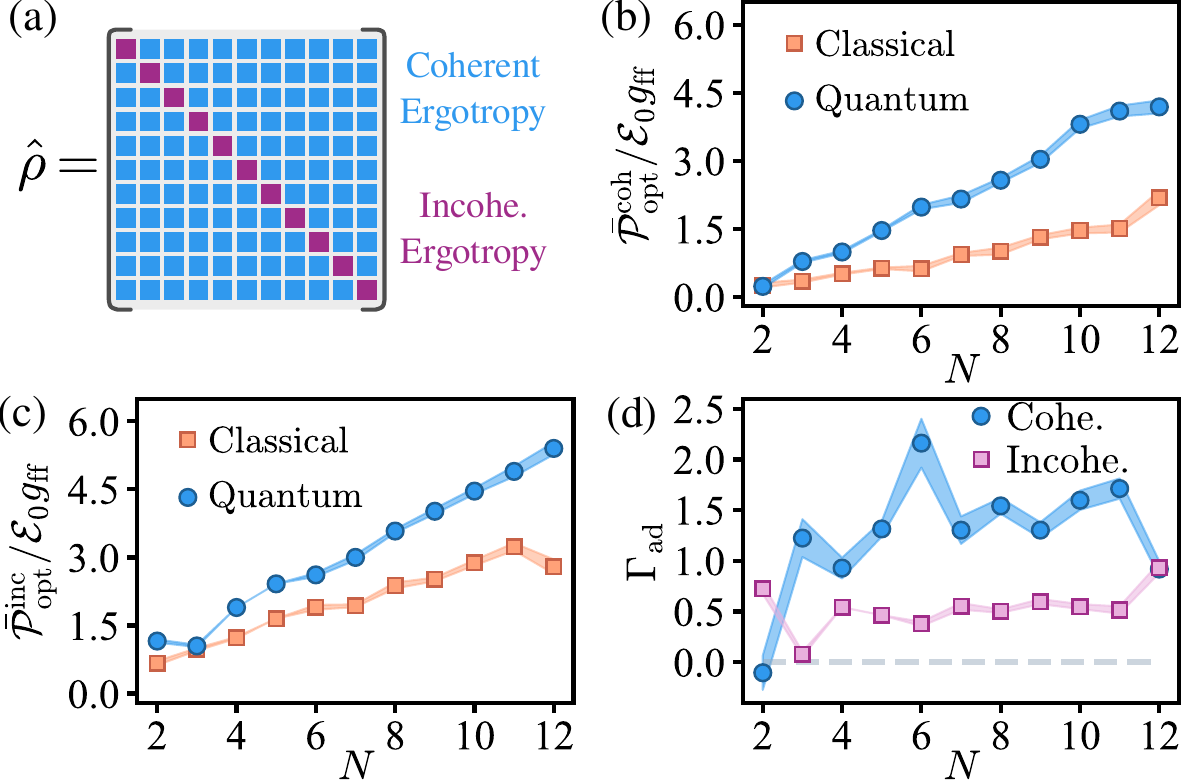}
		\caption{(a) Schematic representation of density matrix coherences and populations, and their connection to coherent and incoherent components of the total ergotropy. (b-c) Behavior of optimal coherent (b) and incoherent (c) ergotropy power as a function of the number of cells $N$, for classical and quantum charging. (d) Coherent and incoherent power advantage parameter as a function of $N$. \RefA{All datasets are based on experimental data, with their error bars shown as shaded regions.}}
		\label{Fig:Incoherent_Ergotropy}
	\end{figure}
	
	This strong bunching of excitations witnessed through the analysis of the $g^{(2)}$ is the positive effect on the charging of the QB, as it means that the QB is charged through the injection of two-photon correlated excitation in the system.
	
	\RefB{\emph{Quantum features.--}} \RefB{Evidence of the quantum characteristics of our QB is done by computing coherent and incoherent ergotropy in our system, and the emergence of entanglement during the charging process~\cite{Francica:20,Shi:22} (see e.g. End Matter section).} As introduced in Ref.~\cite{Francica:20}, the total ergotropy of the system can be divided into its coherent and incoherent part, as depicted in Fig.~\SubFig{Fig:Incoherent_Ergotropy}{a}, corresponding to the coherence and population in the energy basis, respectively. The incoherent contribution of the ergotropy $\Ecal_{\mathrm{inco}} = \Ecal(\hat{\delta}_{\rho})$ is obtained with respect to the fully-dephased density matrix, $\hat{\delta}_{\rho}$, and the coherent part $\Ecal_{\mathrm{cohe}} = \Ecal(\hat{\rho}_{\mathrm{c}})$ takes into account the energy extracted from coherences of the density matrix, denoted by $\hat{\rho}_{\mathrm{c}}$. 
	
	In our experiment, while the extraction of the coherent ergotropy demands the full quantum tomography, whereas the incoherent part of the ergotropy can be obtained from joint measurement in the  energy basis. For this reason, we experimentally determine $\Ecal(\hat{\delta}_{\rho})$, and use it to indirectly quantify the coherent ergotropy through the ergotropy relation $\Ecal_{\mathrm{cohe}} = \Ecal_{\mathrm{total}} - \Ecal_{\mathrm{inco}}$. The time evolution of both coherent and incoherent ergotropies is used to compute their respective time-averaged powers, and then we extract their optimal values, $\Pcal^{\mathrm{cohe}}_{\mathrm{opt}}$ and $\Pcal^{\mathrm{inco}}_{\mathrm{opt}}$, as a function of the size $N$. The result is shown in Fig.~\SubFig{Fig:Incoherent_Ergotropy}{b} and Fig.~\SubFig{Fig:Incoherent_Ergotropy}{c} for coherent and incoherent power, respectively. Consequently, we are also able to quantify the advantage quantum charging parameter related to coherent and incoherent ergotropies as shown in Fig.~\SubFig{Fig:Incoherent_Ergotropy}{d}.
	
	The possibility of scalability of our model allows us to empirically observe some features that classical batteries are not accessible for work extraction of single-cell batteries~\cite{Alexia:20,Niu:24}. We observe a significant enhancement of the coherent power for $N>2$, compared with the counterpart for QBs with $N=2$ cells. We attribute this unexpected feature as a consequence of the anti-blockade effect, as witnessed by $g^{(2)}$, in the quantum charging. In fact, for the case $N=2$, the coherences due to transitions into the single-excitation subspace are reduced for the quantum charging---the transitions $\ket{\downarrow\downarrow} \rightleftarrows \ket{\uparrow\uparrow}$ dominate the evolution. However, for $N > 2$ \RefC{the QCA} observed in our experiment arises from the combined contributions of both coherent and incoherent charging power.

	\emph{Conclusion.--} We have reported clear theoretical and experimental evidence \RefC{of scalable QCA} in solid-state QBs, with substantial enhancement observed for battery sizes up to 12 cells. When imposing a fair energy-demanding scenario, the figures of merit demonstrate a quantum charging outperforms its classical counterpart for all cases considered in this work. Furthermore, QCA performance improves with increasing battery size, highlighting its scalability. We have also found evidence that the performance of the charging is mainly related to an excitation-facilitating mechanism similar to the anti-blockade phenomena, as locally each pair of cells are driven by two-photon transitions.	\RefB{We argue that  the QCA of the quantum charging protocol arises from the interplay of collective nonlinear dynamics and intrinsic quantum resources, such as quantum coherence and entanglement.} The nearest-neighbor aspect of the charging Hamiltonian opens a broad avenue for further development of interacting-efficient models for QBs. In particular, our findings explicitly address the role of short-range and pairwise interactions in enhancing charging efficiency, which makes them highly compatible with current experimental platforms.
	
	\begin{acknowledgments}
	\textit{Acknowledgments.--} This work is supported by the National Natural Science Foundation of China (12574550, 11934010, 12205137, 12004167, 12404582), the Key-Area Research and Development Program of Guangdong Province (Grants No. 2018B030326001),  the Quantum Science and Technology-National Science and Technology Major Project (Grant No. 2021ZD0301703). M.S.S. is supported by Conselho Nacional de Desenvolvimento Cient\'{\i}fico e Tecnol\'ogico (CNPq) (grant number 303836/2024-5).
	This research is also supported in part by Coordena\c{c}\~ao de Aperfei\c{c}oamento de Pessoal de N\'{\i}vel Superior (CAPES) (Finance Code 001) 
	and by the Brazilian National Institute for Science and Technology of Quantum Information (INCT-IQ). ACS is supported by the Comunidad de Madrid through the program Ayudas de Atracción de Talento Investigador ``César Nombela", under Grant No. 2024-T1/COM-31530 (Project SWiQL).
	\end{acknowledgments}

	\textbf{Data Availability}. The Python notebook used to generate all theoretical data in the main text and Supplementary Material is available on GitHub~\cite{GitHub_SQB}. The experimental data supporting the findings of this study can be obtained upon reasonable request from the corresponding authors.

	
	%

	
	\section*{End Matter}

	\subsection{Quantum charging bounds}
	
	In this section, we present a complementary discussion to the charging power scaling considered in the main text, which is related to the quantum charging bounds introduced in Ref.~\cite{Kim:22}. Now, we show that our experiments are in agreement with such bounds. To this end, first, we consider the behavior of the \textit{instantaneous} charging power, rather than the average power previously considered.
	
	This analysis can be done through the power operator, as defined in~\cite{Santos:20c}, used to derive bounds on the charging performance of both classical and quantum charging processes~\cite{Kim:22}. In our case, the power operator for the classical/quantum charging is given by
	\begin{align}
		\hat{P}_{\mathrm{cl}/\mathrm{qu}} = \frac{1}{\hbar} [H_{0},V_{\mathrm{cl}/\mathrm{qu}}]  ,
	\end{align}
	which provides the charging bound (according to Corollary 1 in Ref.~\cite{Kim:22})
	\begin{align}
		| P_{\mathrm{cl}/\mathrm{qu}} (t) | \leq ||\hat{P}_{\mathrm{cl}/\mathrm{qu}}|| \le \frac{k_{\mathrm{cl}/\mathrm{qu}}}{2} \omega_0 v^{ \text{dv}}_{\mathrm{cl}/\mathrm{qu}} , \label{Eq:Ineq}
	\end{align}
	where $k_{\mathrm{cl}}$ and $k_{\mathrm{qu}}$ are the maximum number of cells that are collectively charged in the classical and quantum charging, respectively. We can use that $k_{\mathrm{cl}}=1$ for the classical charging, and $k_{\mathrm{qu}}=2$ for our quantum charging Hamiltonian, as it evolves only pairwise interactions, which yields the bounds
	\begin{align}
		r_{\mathrm{cl}}(t,N) = \frac{| P_{\mathrm{cl}} (t) |}{ \omega_0 v^{ \text{dv}}_{\mathrm{cl}}} \leq \frac{1}{2} ,
		\quad  
		r_{\mathrm{qu}}(t,N) =  \frac{| P_{\mathrm{qu}} (t) |}{ \omega_0 v^{ \text{dv}}_{\mathrm{qu}}} \leq 1 . \label{Eq:Bounds}
	\end{align}
	
\begin{figure}[t]
		\includegraphics[width=\linewidth]{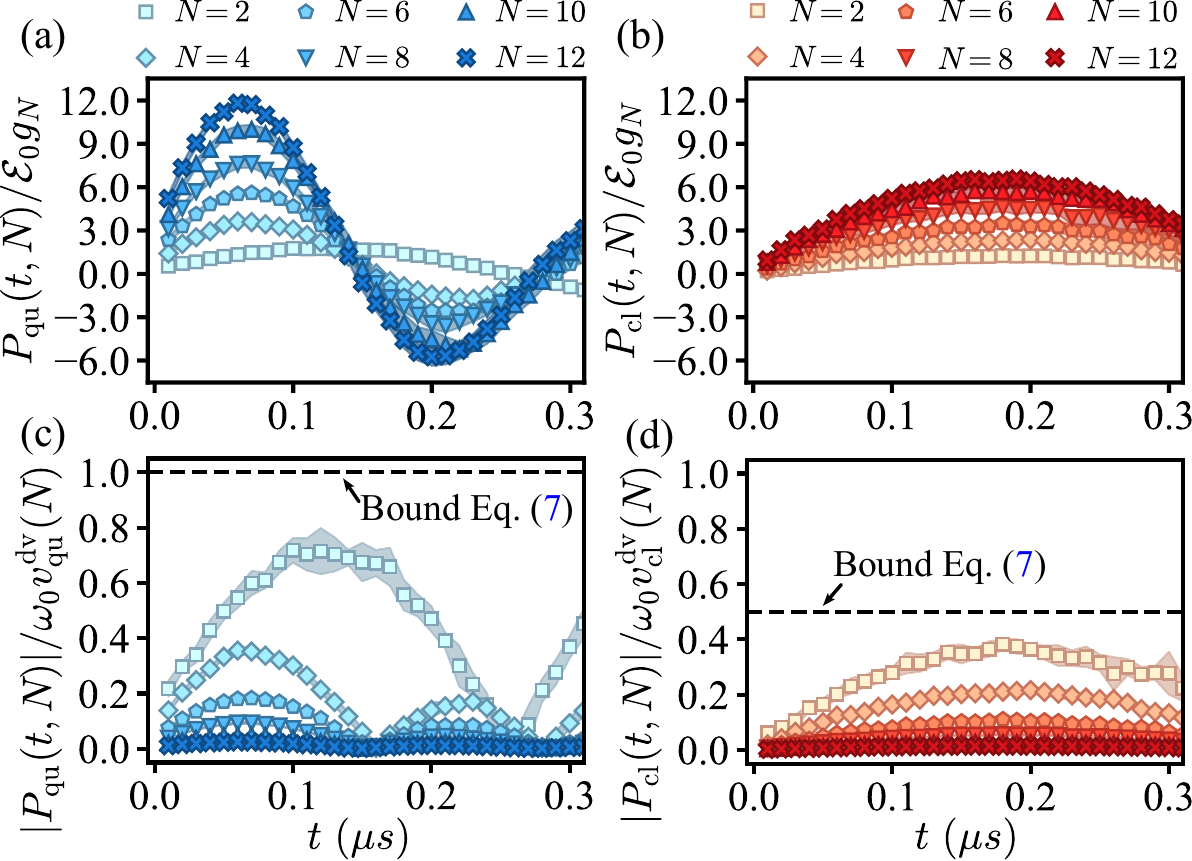}
		\caption{Experimental data for instantaneous power as a function of time for different values of $N$, for (a) quantum and (b) classical charging. Quantum charging bound for (c) quantum and (d) classical charging as defined from Eq.~\eqref{Eq:Bounds}. Experimental parameters are similar to Fig.~\ref{Fig:QCA_scaling}. Error bars indicating the statistical uncertainty are shown as shaded regions.}
		\label{Fig:Bounds}
	\end{figure}

	As these quantities are time-independent, it means that classical and quantum charging do not violate such bounds, which is verified in our experiment. From our experimental data we calculate the instantaneous power as $P(t) = \Delta\mathcal{E}(t) / \Delta t$, where $\Delta\mathcal{E}(t) = \mathcal{E}(t+\Delta t) - \mathcal{E}(t)$ and $\Delta t = 20 \text{ ns}$. In Fig.~\ref{Fig:Bounds} we show the instantaneous charging power for the quantum, see Fig.~\SubFig{Fig:Bounds}{a}, and classical charging in Fig.~\SubFig{Fig:Bounds}{b}, where we can see the high performance of the quantum charging protocol in the interval $t \leq \tau_{\mathrm{max}}$. Also, it is possible to see that the instantaneous power assumes negative values, which is expected due to the spontaneous coherent discharging process~\cite{Santos:20c,Luiz:21}, as our charging is not adiabatic and therefore unstable. In conclusion, the maximum of the instantaneous power scales with the number of cells $N$ in a similar manner to the average power shown in the main text.
	
	Complementarily, in Figs.~\SubFig{Fig:Bounds}{c} and~\SubFig{Fig:Bounds}{d}, respectively, time-dependent behavior of the quantities $r_{\mathrm{qu}}(t,N)$ and $r_{\mathrm{cl}}(t,N)$, defined in Eq.~\eqref{Eq:Bounds}, are shown. One sees that by increasing the number of cells in the battery, the coefficients $r_{\mathrm{cl}/\mathrm{qu}}(t,N)$ tend to small values. This demonstrates that, as expected, the bounds in Eq.~\eqref{Eq:Bounds} are satisfied for all cases considered in our work.
	
	It is timely to say that the analysis considered in the main text, through the time-averaged power, also satisfies such bounds. In fact, the average power defined in the main text can be written as
	\begin{align}
		\bar{\Pcal}_{\mathrm{cl}/\mathrm{qu}}(\tau) = \frac{1}{\tau} \int_{0}^{\tau} P_{\mathrm{cl}/\mathrm{qu}}(t) dt ,
	\end{align}
	What allows us to use the Cauchy-Schwarz inequality to derive the following inequality
	\begin{align}
		|\bar{\Pcal}_{\mathrm{cl}/\mathrm{qu}}(\tau)| = \left\vert\frac{1}{\tau} \int_{0}^{\tau} P_{\mathrm{cl}/\mathrm{qu}}(t) dt \right\vert \leq \frac{1}{\tau} \int_{0}^{\tau} |P_{\mathrm{cl}/\mathrm{qu}}(t)| dt ,
	\end{align}
	and therefore we can invoke the time-independent inequality in Eq.~\eqref{Eq:Ineq}
	\begin{align}
		|\bar{\Pcal}_{\mathrm{cl}/\mathrm{qu}}(\tau)| \leq \frac{k_{\mathrm{cl}/\mathrm{qu}}}{2} \omega_0 v^{ \text{dv}}_{\mathrm{cl}/\mathrm{qu}} .
	\end{align}
	
	\RefCnew{As a final remark on the bounds $r_{\mathrm{cl}}$ and $r_{\mathrm{qu}}$ in Eq.~\eqref{Eq:Bounds}, the scaling behaviors observed in Figs.~\SubFig{Fig:Bounds}{c} and~\SubFig{Fig:Bounds}{d} suggest a dependence of the bounds on the number of cells $N$ in the battery, which contrast with the increasing behavior of the maximum power in Figs.~\SubFig{Fig:Bounds}{a} and~\SubFig{Fig:Bounds}{b}. We understand this result as a consequence of the broad tightness of the bounds developed so far, as this decreasing behavior of the bound dependent on $N$ is not predicted by the $k$-partition parameters $k_{\mathrm{cl}/\mathrm{qu}}$. Our results therefore call for the development of tighter theoretical bounds for quantum batteries.} 

\begin{figure}[t]
	\includegraphics[width=\linewidth]{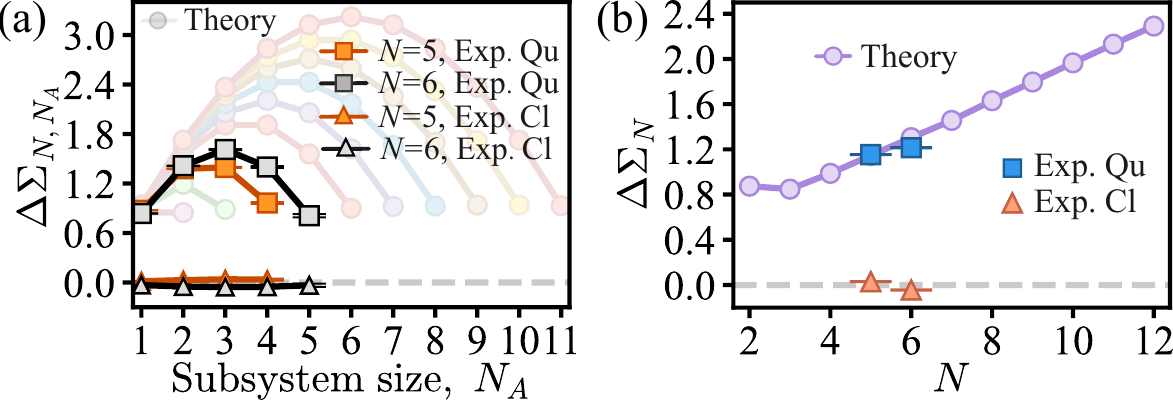}
	\caption{(a) Entropy growth as a function of the size of the subsystem $N_{A}$, for classical (triangles) and quantum charging (square) for $N=\{5,6\}$, with the theoretical curves (circles) shown in the background for comparison. (b) Scaling of the average entropy as a function of the number of qubits in the battery, $N$. In both (a) and (b), the dashed gray line is the expected entropy growth of the classical charging (no entanglement).}
	\label{Fig:Entanglement}
\end{figure}	

\subsection{Quantum correlations in the battery}

We also measure the emergence of entanglement during the charging process witnessed by the second order Rényi entropy~\cite{Horodecki:96PLA,Horodecki:96,Horodecki:09}. We consider the $N$ qubits system partitioned into two subsystems $A$ and $B$, with $N_A$ and $N_B$ qubits, respectively, and we use the second order Rényi entropy, $S_{N,A_i,N_A} = - \log \big[\mathrm{tr}(\rho_{A_i}^2)\big]$. $\rho_{A_i}$ is the density matrix of the reduced system $\{A_i\}$, which contains all possible choices of the subsystems with $N_A$ qubits. Given the multipartite aspect of the system, we compute the bipartitions averaged entropy growth during the charging process as (see Ref.~\cite{SM} for further details)
\begin{equation}
	\Delta \Sigma_{N,N_A} = \frac{1}{\mathrm{dim}\{A_i\}} \sum\nolimits_{\{A_i\}} \left[S_{N,A_i,N_A}(\Delta t) - S_{N,A_i,N_A}(0)\right] .
\end{equation}
for a time interval $\Delta t \approx 0.107 \, \mu\text{s}$, which is approximately the time required to reach maximum charging power---see e.g. Fig.~\ref{Fig:QCA_scaling}. The $\mathrm{dim}\{A_i\}$ is the number of distinct bipartitions of size $N_A$. Also, we compute the scaling of the average entropy increases as a function of the battery size, $N$. For this purpose, we compute the entropies for all possible partition sizes $N_A$, and define the average entropy growth obtained by equally weighting all bipartition sizes of a system with $N$ qubits
\begin{equation}
	\Delta \Sigma_{N} = \frac{1}{N-1} \sum\nolimits_{N_A = 1}^{N-1} \Delta \Sigma_{N,N_A} .
\end{equation}

The measurements of the entropy were experimentally performed for battery sizes $N=5$ and $N=6$, but we included theoretical data to complete the analysis for battery sizes from $N=2$ up to $N=12$. The result is shown in Fig.~\ref{Fig:Entanglement}. When compared to the classical counterpart, the bipartitions averaged entropy growth shown in Fig.~\SubFig{Fig:Entanglement}{a}, and the total average entropy in Fig.~\SubFig{Fig:Entanglement}{b} are clear evidence of the emergence of entanglement during the charging process~\cite{Horodecki:96PLA,Horodecki:96,Horodecki:09}.

\newpage
\onecolumngrid

\newpage

\begin{center}
	{\large{ {\bf Supplemental Material for: \\ \Title}}}
	
	\vskip0.5\baselineskip{
		Chang-Kang Hu,$^{1,2,{\color{blue}\ast}}$ Chilong Liu,$^{1,2,3{\color{blue}\ast}}$ Jingchao Zhao,$^{1,2,3}$ Liuzhu Zhong,$^{1,2,3}$ 
		Yuxuan Zhou,$^{1,2}$ Mingze Liu,$^{1,2,3}$ \\Haolan Yuan,$^{1,2,3}$ Yongchang Lin,$^{1,2,3}$ Yue Xu,$^{1,2,3}$ Guantian Hu,$^{1,2,3}$ 
		Guixu Xie,$^{1,2,3}$ Zixing Liu,$^{1,2}$ Ruiyang Zhou,$^{1,2,3}$ \\Yougui Ri,$^{1,2,3}$ Wenxuan Zhang,$^{1,2,3}$ Ruicheng Deng,$^{1,2,3}$ 
		Andreia Saguia~\orcidlink{0000-0003-0403-4358},$^{4}$ Xiayu Linpeng,$^{1,2}$ Marcelo S. Sarandy~\orcidlink{0000-0003-0910-4407},$^{4}$ \\
		Song Liu,$^{1,2,5}$ Alan C. Santos~\orcidlink{0000-0002-6989-7958},$^{6,\color{blue}\dagger}$ Dian Tan,$^{1,2,5,\color{blue}\ddagger}$ and Dapeng Yu$^{1,2,5}$}
	
	\vskip0.5\baselineskip{
		{\it$^{1}$ \szkl}\\
		{\it $^{2}$\gdpkl}
		\\
		{\it $^{3}$\siqse}
		\\
		{\it $^{4}$\UFF }
		\\
		{\it $^{5}$\HFNL}
		\\
		{\it $^{6}$\CSIC}
	}
	
	\vskip0.5\baselineskip{$^{\color{blue}\ast}$These authors contributed equally to this work.\\
		$^{\color{blue}\dagger}$ac\_santos@iff.csic.es ~~~ $^{\color{blue}\ddagger}$tandian@iqasz.cn}\\
	\vspace{0.5cm}
\end{center}

\appendix

\setcounter{equation}{0}
\setcounter{figure}{0}
\setcounter{table}{0}

\renewcommand{\theequation}{S\arabic{equation}}
\renewcommand{\thefigure}{S\arabic{figure}}
\renewcommand{\thesection}{S\arabic{section}}
\renewcommand{\thesubsection}{S\arabic{section}-\arabic{subsection}}

\twocolumngrid

\tableofcontents
\section{DEVICE AND EXPERIMENTAL SETUP}
\subsection{Superconducting quantum processor}

\begin{figure*}[t]
	\centering
	\includegraphics[width=0.85\linewidth]{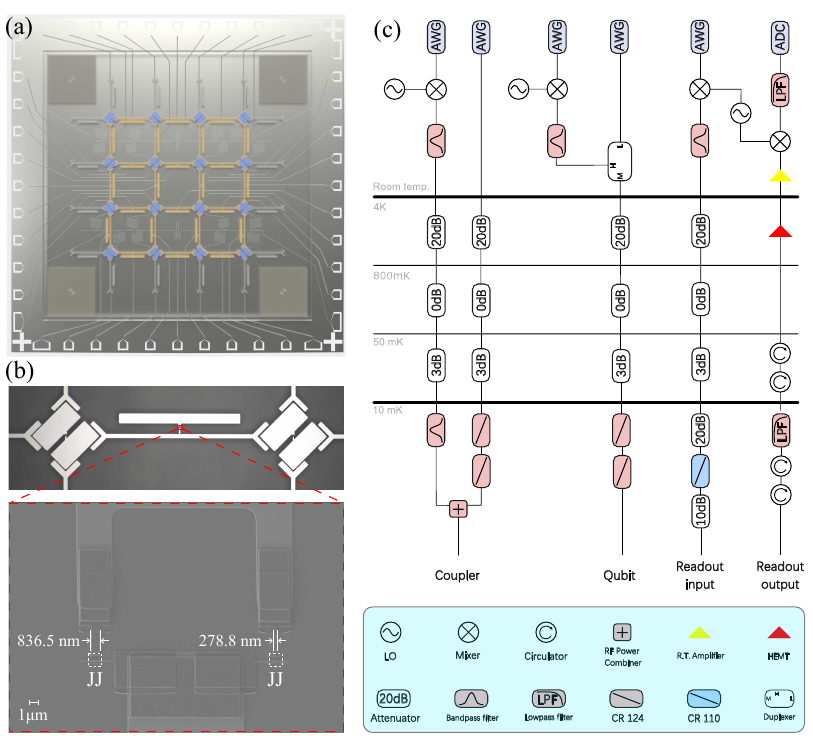}
	\caption{\textbf{Experimental setup.} \textbf{(a)} The 16-qubit
		lattice-structured superconducting quantum processor. \textbf{(b)} Schematic diagram illustrating the experimental wiring, showing the signal lines from the room-temperature control electronics down to the sample stage within the dilution refrigerator.}
	\label{fig: DeviceWiring}
\end{figure*}

Our experiments utilize a superconducting quantum processor comprising of 16 qubits arranged in a 4x4 lattice, as shown in Fig.~\SubFig{fig: DeviceWiring}{a}. This quantum processor is fabricated using a 3D flip-chip packaging technique, forming a two-layer structure, separated by a gap of approximately 9 $\mu$m, supported by patterned photoresist structures. The bottom layer contains the readout resonators, filters, and control lines, while the top layer hosts the tunable-frequency qubits and couplers. 
Notably, these control lines combine XY and Z signals, enabling arbitrary single-qubit gate operations and the precise control of magnetic flux signals required for dynamic frequency tuning of both qubits and couplers.
To minimize the crosstalk, we incorporate air bridges in the resonators and Purcell filters. 
Additionally, we employed a cover bridge design for the control lines, where a metallic layer is deposited above the control lines to confine microwave signal propagation and thereby mitigate crosstalk.

\subsection{Wiring and electronics}
In our experiments, the superconducting quantum processor is mounted on the mixing chamber plate of a Bluefors LD400 dilution refrigerator, operating at a base temperature of approximately 10 mK. 
The experimental wiring schematic is depicted in Fig.~\SubFig{fig: DeviceWiring}{b}. High-frequency microwave control signals are generated using a standard heterodyne upconversion technique. An arbitrary waveform generator (AWG) with a sampling rate of 2.4 GSa/s produces an intermediate frequency (IF) waveform, typically centered around 600 MHz. This IF signal is mixed with a continuous-wave signal from a local oscillator (LO). The resulting signal is then passed through a band-pass filter to select the desired sideband, while suppressing undesired components like LO leakage and image frequencies, thereby ensuring clean control signals. 
This method is employed to generate all required high-frequency signals, including the microwave pulses for qubit manipulation, the probe signals for resonator readout, and the parametric drive signals essential for double excitation dynamics investigated in this work. The generation of low-frequency signals for magnetic flux control is more straightforward, produced directly by separate AWG channels, without the need for frequency mixing.
The readout signal, transmitted through the chip, is routed through a series of cryogenic circulators to ensure robust isolation from the unwanted noise, which is then amplified by a High-Electron-Mobility Transistor (HEMT) amplifier at the 4K stage of the refrigerator. Following additional amplification at room temperature, the signal is down-converted to an intermediate frequency and digitized by an Analog-to-Digital Converter (ADC), enabling the discrimination of the final quantum states.

For standard qubit control, a room-temperature diplexer combines the high-frequency microwave signal and the low-frequency flux bias signal, which is then applied to the qubit. 
For the parametric drive used in double excitation dynamics, however, the wiring is different. This process is relatively slow and requires higher microwave power, so the control line for the parametric drive bypasses the standard cryogenic attenuators and infrared filters (CR124) to avoid excessive signal attenuation. To mitigate thermal noise entering through this less-attenuated line, a band-pass filter centered at the drive frequency is inserted, which transmits the drive signal while rejecting out-of-band thermal noise that could degrade qubit performance.
Additionally, the parametric drive requires a DC flux signal to bias the coupler to its optimal operating point. A separate low-frequency control line for flux bias signal is combined with the high-frequency drive signal using a second combiner before being applied to the coupler.

\subsection{Chip parameters}

To evaluate the performance of the superconducting quantum processor used in our experiments, we conducted a series of characterization experiments. Key device parameters, including qubit frequencies, coherence times, and readout fidelities, were measured, with results summarized in Table \ref{table:Device}. For the qubits utilized in the experiments, we observed average energy relaxation times ($T_1$) of $32.2~\mu$s and average echo coherence times ($T_2^{\text{echo}}$) of $4.6~\mu$s. The average single-shot readout fidelities were determined to be $F_{|0\rangle} = 92.2\%$ for the ground state ($|0\rangle$) and $F_{|1\rangle} = 82.1\%$ for the excited state ($|1\rangle$). The ground state readout fidelity ($F_{|0\rangle}$) is mainly limited by imperfections in quantum state initialization in our experiment.

\begin{table*}[htp]
	
	\renewcommand
	\arraystretch{2.5}
	\centering
	\begin{threeparttable}
		\begin{tabular}{|c|c|c|c|c|c|c|c|c|c|c|c|c|}
			
			\hline
			\hline

			Qubit$^{\rm a}$                           & $\text{Q}_1$  & $\text{Q}_2$  & $\text{Q}_3$  & $\text{Q}_4$  & $\text{Q}_5$  & $\text{Q}_6$  & $\text{Q}_7$  & $\text{Q}_8$  & $\text{Q}_9$   & $\text{Q}_{10}$ & $\text{Q}_{11}$ & $\text{Q}_{12}$ \\
			\hline
			Idling frequency (GHz)                    & 4.575  & 4.249  & 4.412  & 3.897  & 4.292  & 4.445  & 3.999  & 4.303  & 3.897   & 4.242    & 4.007    & 4.408    \\
			Anharmonicity (MHz)                       & -199.96& -200.26& -200.13& -200.48& -200.53& -200.56& -200.42& -200.56& -199.89 & -200.36  & -200.76  & -199.99  \\
			Resonator frequency (GHz)                 & 6.271  & 6.512  & 6.316  & 6.492  & 6.248  & 6.522  & 6.285  & 6.423  & 6.305   & 6.482    & 6.389    & 6.542    \\
			Resonator linewidth (MHz)                 & 0.77   & 0.90   & 0.82	 & 0.54   & 0.80  	 & 0.82   & 0.63   & 0.98   & 0.80    & 0.55     & 0.89     &0.82      \\
			
			Dispersive shift of $|1\rangle$ (MHz)     & 1.18   & 0.84   & 1.07   & 0.59   & 0.83   & 0.80   & 0.64   & 0.80   & 0.61    & 0.68     & 0.60     & 0.69     \\
			Readout fidelity of $|0\rangle$ (\%)      & 93.6   & 95.2	& 95.5   & 90.4   & 94.2   & 90.1	  & 85.9   & 90.7   & 92.9    & 92.4     & 91.6     & 93.7     \\
			Readout fidelity of $|1\rangle$ (\%)      & 85.1   & 84.6	& 84.1   & 87.0   & 85.8   & 81.7   & 78.6   & 76.5   & 79.2    & 78.7     & 78.8     & 85.1     \\
			Relaxation time of $|1\rangle$  ($\mu s$) & 28.7   & 39.9   & 37.0   & 54.1   & 17.5   & 19.5   & 36.9   & 32.8   & 37.0    & 30.6     & 23.5     & 29.0     \\
			Ramsey decay time ($\mu s$)               & 8.1    & 1.4    & 1.7    & 1.6    & 2.3    & 1.9    & 1.5    & 1.6    & 1.2     & 1.5      & 2.4      &3.5       \\
			Spin echo decay time ($\mu s$)            & 14.3   & 3.9    & 3.4    & 3.1    & 3.8    & 3.1    & 3.3    & 3.0    & 2.7     & 2.4      & 4.1      &7.9       \\

			\hline
			\hline
		\end{tabular}
		
		\begin{tablenotes}
			\footnotesize
			
			\item[a]  The table presents the measured parameters when the couplers were idle near the ZZ coupling turn-off point.
			
		\end{tablenotes}
	\end{threeparttable}
	\caption{\textbf{Device parameters}.}
	\label{table:Device}
\end{table*}

\subsection{Readout crosstalk and correction}

In our experiment, accurately determining the quantum state is critical for studying the dynamics of superconducting qubits. However, the fidelity of this readout process is limited by various experimental imperfections. The primary sources of readout error include: firstly, state transitions during the finite measurement duration, induced from energy relaxation ($T_1$ decay: $|1\rangle \to |0\rangle$) and thermal excitation ($|0\rangle \to |1\rangle$); and secondly, errors in state discrimination, which occur when the measurement signals for the $|0\rangle$ and $|1\rangle$ states overlap, leading to misclassification.
To correct for measurement errors, we can use the model $\xi_{\text{ideal}} = M \xi_{\text{noisy}}$, where $\xi_{\text{ideal}}$ denotes the vector of ideal qubit populations, $\xi_{\text{noisy}}$ the vector of measured qubit populations, and $M$ the response matrix that maps the noisy to the ideal probability distribution.

\begin{figure*}[t]
	\centering
	\includegraphics[width=0.85\linewidth]{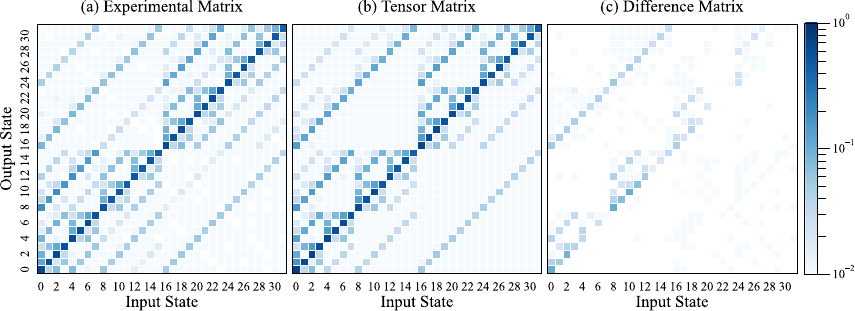}	
	\caption{\textbf{5-Qubit Readout Matrices.} \textbf{(a)} Experimentally measured matrix. \textbf{(b)} Readout matrix derived from single-qubit readout matrix, neglecting the inter-qubit crosstalk. \RefA{(c) Shows the absolute value of the difference between the Experimental Matrix and Tensor Matrix.}}
	\label{fig: Readout_matrix}
\end{figure*}

Here, to illustrate our readout correction procedure, we focus on a five-qubit system. This method is consistently applied across all systems with 2 to 12 qubits. For this five-qubit system, each of the $2^5=32$ computational basis states was prepared, followed by joint measurements. Figure~\SubFig{fig: Readout_matrix}{a} shows the resulting $32 \times 32$ readout matrix, where each element $P(i|j)$ represents the probability of measuring the basis state $|i\rangle$ when the state $|j\rangle$ was prepared. For comparison, Figure~\SubFig{fig: Readout_matrix}{b} shows a theoretical readout matrix which is constructed assuming independent single-qubit readout errors and ignoring any readout crosstalk between qubits. 

We observe good agreement between the measured multi-qubit readout matrix (c.f. Fig.~\SubFig{fig: Readout_matrix}{a}) and the theoretical readout matrix (c.f. Fig.~\SubFig{fig: Readout_matrix}{b}) which provides that inter-qubit readout crosstalk is negligible in this system. This result supports the use  of error mitigation strategies that treat individual qubit readout errors independently. Therefore, we applied the readout correction techniques from the Mthree (Matrix-Free Measurement Mitigation)~\cite{Nation:21} software package throughout our data analysis.

\section{PARAMETRIC DRIVING SCHEME} 

\subsection{Parametric driving theory}

To implement the bSWAP operations described by Eq.~(1) in the main text, we employ a parametric driving scheme~\cite{Marco:17}. To show how to implement this dynamics, we first consider a simple system which is composed of two qubits (Q1, Q2) coupled via a tunable coupler (C). The Hamiltonian describing this system in the lab frame is given by:
\begin{equation}
	H = -\sum_{i=1,2} \frac{\omega_i}{2}\sigma_i^z - \frac{\omega_c(\Phi(t))}{2}\sigma_c^z + g_1 \sigma_1^x \sigma_c^x + g_2 \sigma_2^x \sigma_c^x 
\end{equation}
Here, $\sigma_{i(c)}^{z,x}$ are the Pauli operators for the qubits ($i=1,2$) and the coupler ($c$), $\omega_i$ are the qubit frequencies, $g_i$ represent the qubit-coupler coupling strength, and $\omega_c(\Phi(t))$ is the tunable frequency of the coupler, which depends on the applied magnetic flux $\Phi(t)$.

The tunable coupler is designed by employing an asymmetric DC SQUID loop. This asymmetry is characterized by the parameter $d = |E_{J1}-E_{J2}|/(E_{J1}+E_{J2})$, where $E_{J1}$ and $E_{J2}$ are the Josephson energies of the two junctions forming the SQUID. The frequency of this asymmetric coupler, $\omega_c$, can be tuned by the external flux $\Phi$ threading the SQUID loop. Its dependence on flux is approximately given by:
\begin{equation}
	\omega_c(\Phi) \approx \omega_{c,max} \left[ \cos^2\left(\pi \frac{\Phi}{\Phi_0}\right) + d^2 \sin^2\left(\pi \frac{\Phi}{\Phi_0}\right) \right]^{1/4} , 
	\label{SM-Eq:omega}
\end{equation}
where $\Phi_0 = h/(2e)$ is the magnetic flux quantum, and $\omega_{c,max}$ is the maximum frequency achieved at integer flux quanta ($\Phi = n\Phi_0$). The non-zero asymmetry ($d>0$) ensures that the coupler frequency remains finite at half-integer flux quanta, $\omega_c(\Phi_0/2) \approx \omega_{c,max} \sqrt{d}$, enhancing resilience to flux noise at certain operating points compared to a symmetric SQUID ($d=0$).

\begin{figure}[t]
	\centering
	\includegraphics[width=\linewidth]{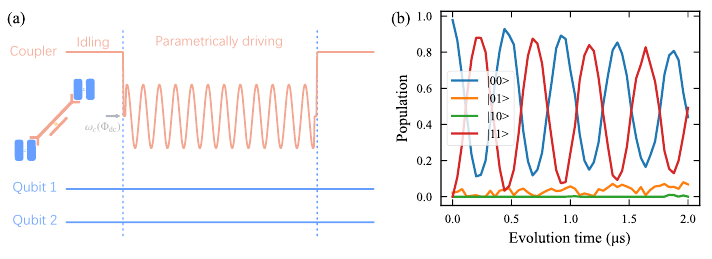}
	\caption{\textbf{Experimental pulse sequence and resulting population dynamics.} \textbf{(a)} Experimental pulse sequence illustrating the sequence of control pulses applied during the experiment. \textbf{(b)} Time evolution of the measured populations, recorded under the experimental protocol depicted in panel (a).}
	\label{fig: bswap_dynamics}
\end{figure}

To mediate a tunable interaction as a bSWAP interaction between the qubits, we employ parametric modulation of the coupler's frequency, as shown in in Fig.~\SubFig{fig: bswap_dynamics}{a}. This is achieved by applying a time-varying component to the magnetic flux around the static bias $\Phi_{\text{dc}}$:
\begin{equation}
	\Phi(t) = \Phi_{\text{dc}} + \delta_\Phi \cos(\omega_\Phi t + \phi_{0})
	\label{eq:Phi_t}
\end{equation}
Here, $\delta_\Phi$ is the modulation amplitude, $\omega_\Phi$ is the modulation frequency, and $\phi_{0}$ is the phase. This flux modulation makes the coupler frequency $\omega_c(\Phi(t))$ time-dependent according to Eq.~\eqref{SM-Eq:omega}.

In our experiment, the modulation frequency $\omega_\Phi$ is set near the sum of the qubit frequencies, i.e., $\omega_\Phi \approx \omega_1 + \omega_2$. This resonant condition activates a parametric interaction between the two neighboring qubits. By transforming into an appropriate interaction picture and applying the Schrieffer-Wolff transformation, one can derive an effective Hamiltonian describing the interaction between Q1 and Q2:
\begin{equation}
	H_{\text{eff}} \approx \Omega_{\text{eff}} (\sigma_1^+ \sigma_2^+ + \sigma_1^- \sigma_2^-)
\end{equation}
This Hamiltonian generates the desired bSWAP dynamics. The strength of this effective interaction, $\Omega_{\text{eff}}$, is given by:
\begin{equation}
	\Omega_{\text{eff}} \approx -\delta_\Phi\frac{g_1 g_2}{4} \left. \frac{\partial \omega_c}{\partial \Phi} \right|_{\Phi_{\text{dc}}} \left( \frac{1}{\Delta_{1,-} \Delta_{2,+}} + \frac{1}{\Delta_{1,+} \Delta_{2,-}} \right)
	\label{SM-Eq:omega_eff}
\end{equation}
where the derivative of the coupler frequency $\partial \omega_c / \partial \Phi$ is evaluated at the DC flux bias $\Phi_{\text{dc}}$. The detuning factors $\Delta_{i,\pm}$ depend on the qubit frequencies, and the coupler frequency at the bias point $\omega_c(\Phi_{\text{dc}})$ (e.g., $\Delta_{i,-} = \omega_i - \omega_c(\Phi_{\text{dc}})$, $\Delta_{i,+} = \omega_i + \omega_c(\Phi_{\text{dc}})$. Equation~\eqref{SM-Eq:omega_eff} reveals how the effective coupling $\Omega_{\text{eff}}$ can be controlled. With the intrinsic parameters ($g_i, \omega_i, \omega_{c,max}, d$) fixed, $\Omega_{\text{eff}}$ is primarily tuned via:
\begin{enumerate}
	\item The DC flux bias $\Phi_{\text{dc}}$: This sets the operating point, determining the coupler's frequency $\omega_c(\Phi_{\text{dc}})$ and its local susceptibility to flux, $\partial \omega_c / \partial \Phi |_{\Phi_{\text{dc}}}$.
	\item The modulation amplitude $\delta_\Phi$: The strength $\Omega_{\text{eff}}$ typically scales with $\delta_\Phi$ (often linearly for small modulation depths).
\end{enumerate}

\subsection{bSWAP implementation}

In our experiments, in the idle configuration (when no interaction is actively driven), the qubits are operated at frequencies $\omega_1/2\pi$ and $\omega_2/2\pi$. The coupler is biased with a static magnetic flux while its idle frequency resides near the point where the static ZZ interaction between Q1 and Q2 is minimized. 
The careful selection of these idle frequencies is vital for minimizing crosstalk and achieving high fidelity. A significant advantage of our method is that the dynamic operation does not need individual qubit control. We only need to apply external magnetic flux control to the coupler, as detailed in Eq.~\eqref{eq:Phi_t}. Figure~\SubFig{fig: bswap_dynamics}{a} shows the pulse sequence for the bSWAP implementation. This strategy allows us to avoid complicated calibration procedures for crosstalk and waveform distortion, thereby ensuring robust experimental outcomes.
Then, we typically choose a bias point $\Phi_{\text{dc}}$ where the coupler frequency $\omega_c(\Phi)$ exhibits a reasonably linear dependence on $\Phi$. 

\begin{figure}[t]
	\centering
	\includegraphics[width=\linewidth]{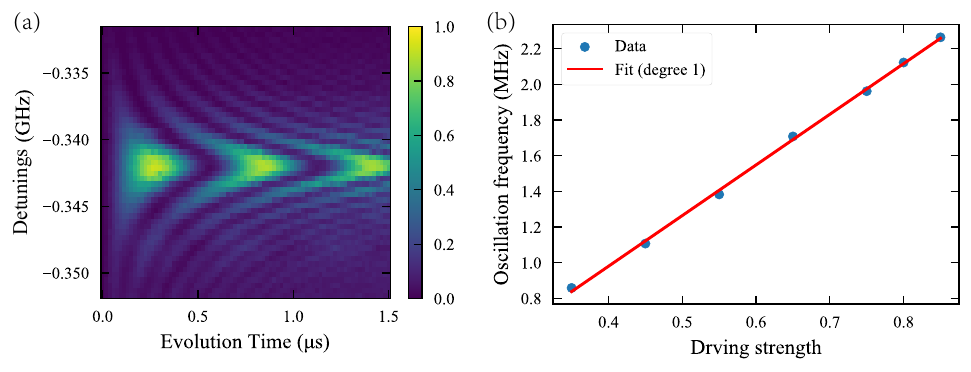}
	\caption{\textbf{Characterization of $|00\rangle \leftrightarrow |11\rangle$ dynamics.} \textbf{(a)} The $|00\rangle \leftrightarrow |11\rangle$ population transfer as a function of the applied drive frequency. This plot highlights the optimal drive frequency for achieving high-fidelity exchange. \textbf{(b)} The oscillation frequency plotted against the drive strength. The data exhibit a linear dependence of the oscillation frequency on the drive amplitude.}
	\label{fig: Bswap}
\end{figure}

Figure~\SubFig{fig: bswap_dynamics}{b} illustrates the time evolution of the populations for the joint measurement results in four computational basis, $|00\rangle$, $|01\rangle$, $|10\rangle$, and $|11\rangle$, under the dynamics. It is evident from this figure that significant population exchange occurs exclusively between the $|00\rangle$ and $|11\rangle$ states. Conversely, when the system is initialized in either the $|01\rangle$ or $|10\rangle$ state, their respective populations remain notably constant throughout the observed evolution, indicating these states are largely decoupled from the driving dynamics.

For optimizing this two-qubit dynamic process, specifically to maximize the fidelity and efficiency of the $|00\rangle \leftrightarrow |11\rangle$ dynamics, careful tuning of the modulation frequency $\omega_\Phi$ is critical in the experiment. The pronounced impact of the modulation frequency on this exchange process is clearly demonstrated in Fig.~\SubFig{fig: Bswap}{a}. Furthermore, within the parameter range explored in our study, we investigated the dependence of the $|00\rangle \leftrightarrow |11\rangle$ exchange strength on the drive amplitude. Our experimental data in Fig.~\SubFig{fig: Bswap}{b} consistently show a linear relationship between the observed exchange strength and the applied drive amplitude over this range.

Building upon this two-qubit implementation, we extend the bSWAP operation to larger systems by sequentially incorporating additional qubits into this one-dimensional quantum battery chain. The same bSWAP-based approach is used to establish nearest-neighbor interactions throughout the chain, enabling scalable control of collective charging across the multi-qubit system.

\section{THEORETICAL DESCRIPTION OF THE QUANTUM CHARGING} \label{Sect:Purity}

\subsection{Pure-state approximation}

\begin{figure}[t]
	\centering
	\includegraphics[width=\linewidth]{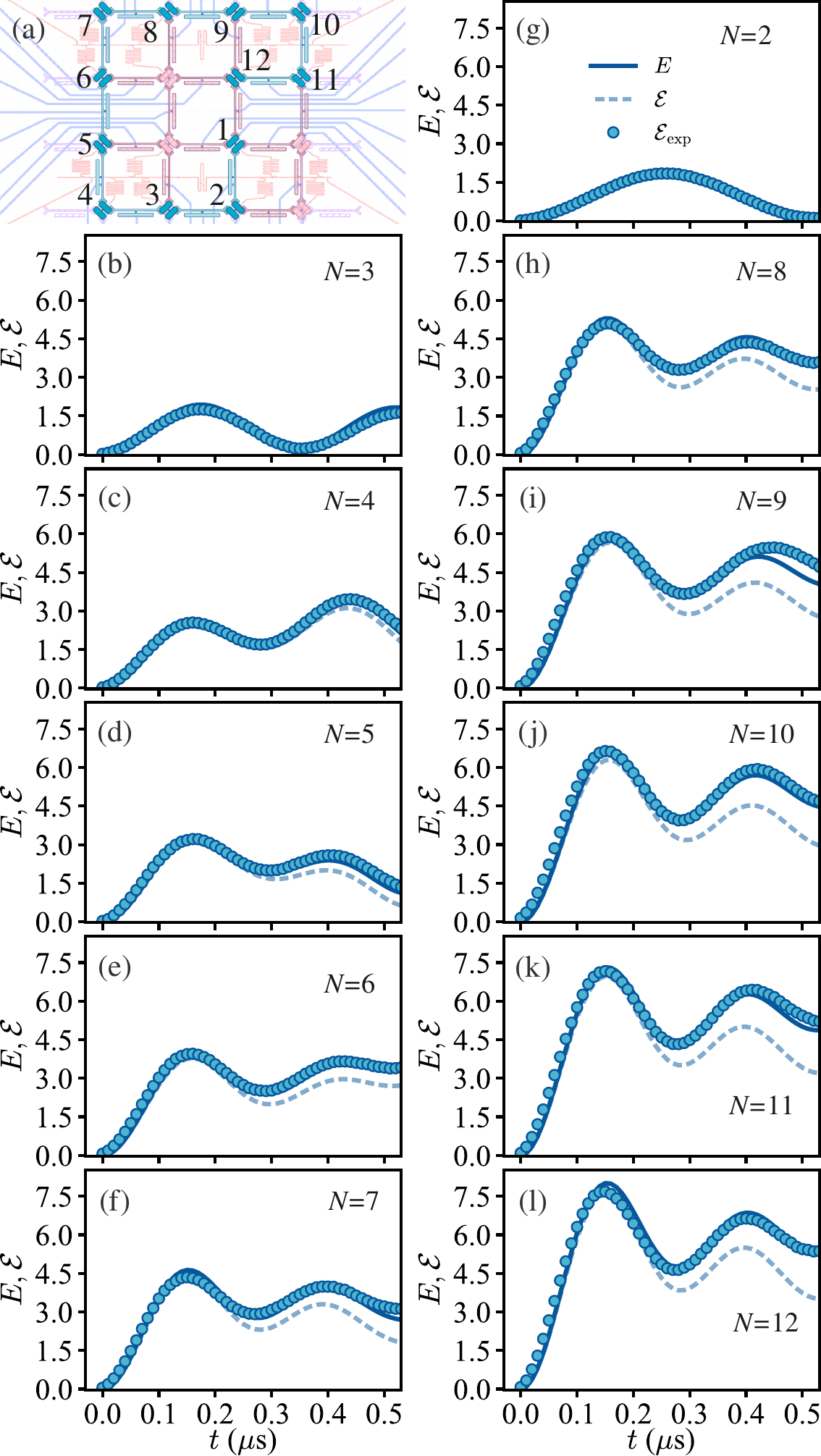}
	\caption{\textbf{Quantum cells mapping and ergotropy.} (a) Sketch shows the qubits used in our experiment (in green), while the inactive qubits remains in their idle point to build an effective $12$-cell quantum battery. (b--l) The graphs show the theoretical ergotropy (solid line), theoretical internal energy (dashed line) and experimental internal energy (dots) for different values of $N$.}
	\label{fig:Ergo_Internal}
\end{figure}

In this section we include all theoretical analysis supported by the experimental data shown in the main text. As a first discussion, let us show that the pure state approximation assumption is valid, given the decay rates and total evolution time required to reach the maximum charging power. Then, we show theoretical fit to other relevant quantities considered in the main text. To this end, we describe the real evolution of the system considering an open-system approach, where the system is driven under the Lindblad master equation
\begin{align}
	\frac{d\hat{\rho}(t)}{dt} = \frac{1}{i\hbar} [ \hat{H}_\mathrm{ch}, \hat{\rho}(t)] +  \Lcal_{\sigma^{-}}[\hat{\rho}(t)] + \Lcal_{\sigma^{+}\sigma^{-}}[\hat{\rho}(t)] , \label{Ap:EqLind}
\end{align}
with $\hat{H}_\mathrm{ch}$ the charging Hamiltonian, and the super-operators $\Lcal_{\sigma^{-}}[\bullet]$ and $\Lcal_{\sigma^{+}\sigma^{-}}[\bullet]$ describing the relaxation and dephasing processes, respectively, obtained from the general Lindblad form
\begin{align}
	\Lcal_{\hat{L}}[\hat{\rho}(t)] = \sum_{n=1}^{N} \Gamma^{\hat{L}}_{n}  \left[ \hat{L}_{n}\hat{\rho}(t)\hat{L}^{\dagger}_{n} - \frac{1}{2} \{ \hat{L}^{\dagger}_{n}\hat{L}_{n}, \hat{\rho}(t)\}\right]
\end{align}
with $\Gamma^{\hat{L}}_{i} $ the decoherence rate of the $n$-th qubit associated to the decoherence channel $\hat{L}_{n}$. 

\begin{figure}[t!]
	\centering
	\includegraphics[width=\linewidth]{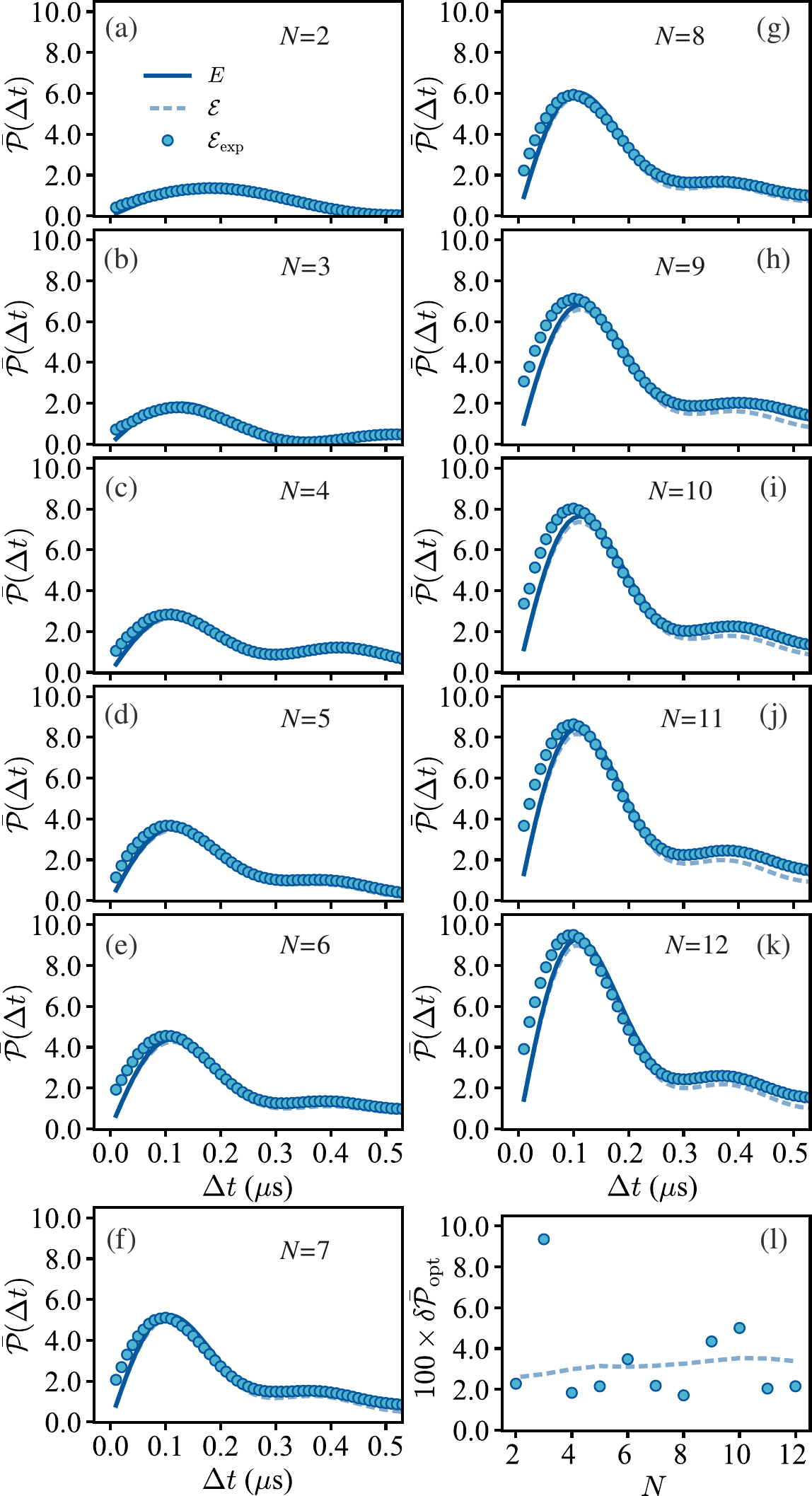}
	\caption{\textbf{Average Powers and its deviations}. (a--k) Average power related to theoretical ergotropy (solid line), theoretical internal energy (dashed line) and experimental internal energy (dots) for different values of $N$. The deviation of the power for theoretical ergotropy and experimental data, with respect to the theoretical internal energy, is shown in (l).}
	\label{fig:AveragePower_Deviation}
\end{figure}

To compare the ergotropy charging for both models, we compute the internal energy $E(t) = \tr(\hat{H}_{0}\hat{\rho}(t))$ and the ergotropy as defined Ref.~\cite{Allahverdyan:04} for ergotropy of arbitrary density matrices. The experimental data corresponds to the internal energy. In Fig.~\ref{fig:Ergo_Internal} we show the behavior of the ergotropy and internal energy obtained form numerical integration of the Eq.~\eqref{Ap:EqLind} (curves), and we compare those results with the experimental data (circle symbol). The theoretical results shown in Fig.~\ref{fig:Ergo_Internal} are obtained from the solution of the Eq.~\eqref{Ap:EqLind} using the experimental parameters provided in Table~\ref{table:Device}. In all result shown here, for long evolution time the decoherence acts on the system and the ergotropy deviates from the internal energy, as expected. However, for short evolution time the ergotropy and internal energy show good agreement.

\begin{figure}[t!]
	\centering
	\includegraphics[width=\linewidth]{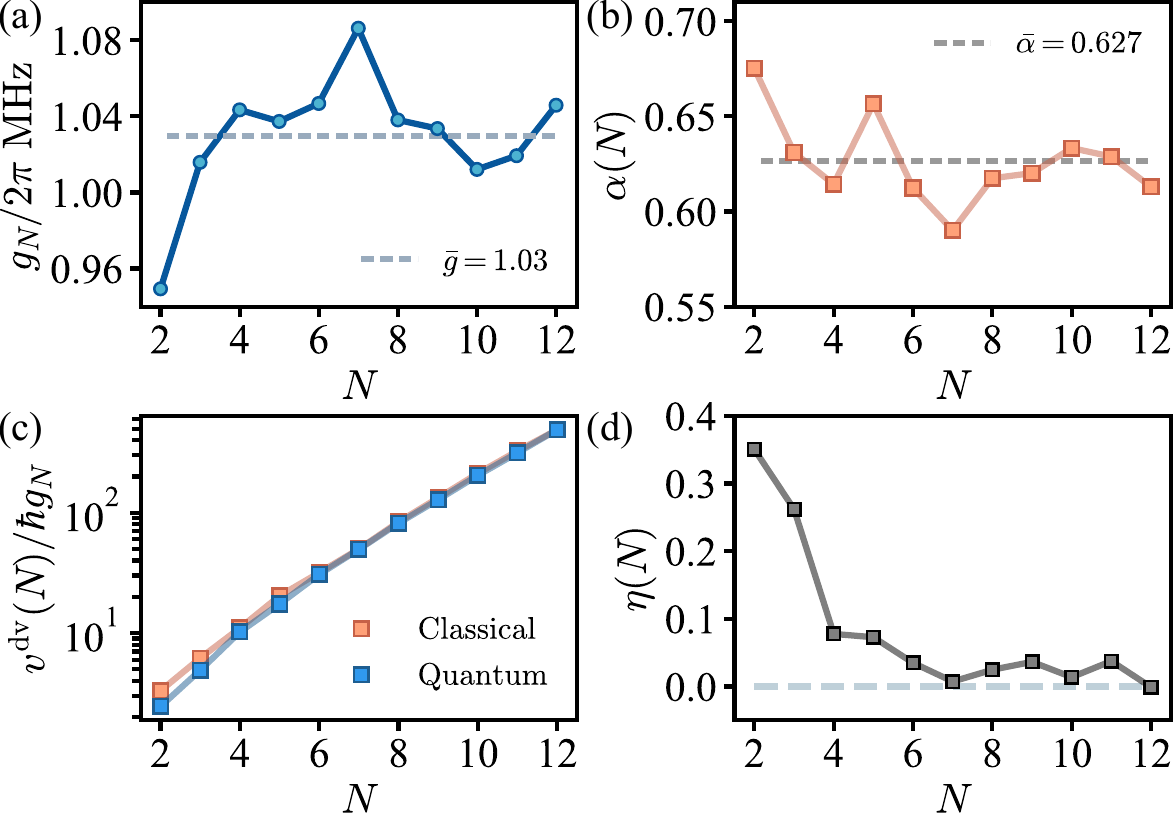}
	\caption{\textbf{Effective couplings and driving potential.} (a) Fitted effective qubit-qubit coupling strength through the master equation~\eqref{Ap:EqLind}. The average value of the coupling $\bar{g} = 1.03 \times 2\pi$~MHz is highlighted through the dashed horizontal line. (b) Coefficient $\alpha(N) = \Omega / g_N$ as function of the number of cells, where we plot its average. (c) Driving potential $v^{\mathrm{dv}}$ as function of $N$, computed as defined in main text. For completeness, in (d) we show the quantum advantage parameter $\eta(N)$. The experimental value for the classical charging field amplitude is $\Omega \approx 0.641\times 2\pi$~MHz.}
	\label{fig:g_measured}
\end{figure}

The value of the effective coupling is obtained by fitting the experimental data, and the result is shown in Fig.~\SubFig{fig:g_measured}{a}, where we also present the values of the other relevant quantities as function of $N$. In particular, we show the values of the parameter $\alpha$ used to set the ratio between the quantum charging coupling and classical charging Rabi frequency, Fig.~\SubFig{fig:g_measured}{b}. These values are used to compute the quantum driving potential $v^\mathrm{dv}$, in Fig.~\SubFig{fig:g_measured}{c}, which is related to the advantage parameter $\eta$ defined in the main text and shown in Fig.~\SubFig{fig:g_measured}{d}. From the data shown in Fig.~\ref{fig:Ergo_Internal} we then compute the average power, and the result is shown in Fig.~\ref{fig:AveragePower_Deviation}. As a complementary discussion to the Fig.~\ref{fig:Ergo_Internal}, we also show the behavior of the average power for the theoretical and experimental internal energy, and the theoretical ergotropy. In particular, we estimate deviations between the three quantities according to the equation\RefA{
	\begin{align}
		\delta \bar{\Pcal}_{\mathrm{opt}} = \frac{|\bar{\Pcal}_{\mathrm{opt}}^{E} - \bar{\Pcal}_{\mathrm{opt}}^{\Ecal}|}{\bar{\Pcal}_{\mathrm{opt}}^{E}}, ~
		\delta \bar{\Pcal}_{\mathrm{opt}}^{{\mathrm{exp}}} = \frac{|\bar{\Pcal}_{\mathrm{opt}}^{E} - \bar{\Pcal}_{\mathrm{opt}}^{E_\mathrm{exp}}|}{\bar{\Pcal}_{\mathrm{opt}}^{E_\mathrm{exp}}} ,
	\end{align}
	which quantifies how much the theoretical optimal power of the ergotropy $\bar{\Pcal}_{\mathrm{opt}}^{\Ecal}$ and the experimental power $\bar{\Pcal}_{\mathrm{opt}}^{E_\mathrm{exp}}$ (obtained from internal energy measurements) deviates of the theoretical power associated to the internal energy $\bar{\Pcal}_{\mathrm{opt}}^{E}$, normalized by $\bar{\Pcal}_{\mathrm{opt}}^{E}$ and $\bar{\Pcal}_{\mathrm{opt}}^{E_\mathrm{exp}}$, respectively}. So, we observe that the maximum deviation of the optimal ergotropy power is around $3.5\%$ of the desired internal energy power $\bar{\Pcal}_{\mathrm{opt}}^{E}$, obtained for perfectly unitary evolutions. Also, apart a significant deviation for the case $N=3$ due to finite control error, the experimental data demonstrate a similar behavior. In particular, we understand that the case $N=3$ does not affect the main message of our work with respect to the QCA. In fact, for $N=3$ the value of the driving ratio $\eta\approx 0.35$ guarantees that we can always change the effective coupling to make the dynamics faster and avoid such a deviation.

\subsection{Going beyond 12-cell quantum batteries}

In this section we discuss the scaling for quantum advantage with larger cells. As result, we show that the scaling of the genuine quantum advantage strongly depend on the parameter $\alpha$, as it has to be considered when characterizing the battery performance.

For simplicity of the numerical analysis, we assume a unitary evolution where no decoherence effect is present in the system. It is done by taking the limit $\Gamma^{\hat{L}}_{i} \rightarrow 0$ in Eq.~\eqref{Ap:EqLind}. So, by initializing the system in the empty ergotropy state of the reference Hamiltonian $\hat{H}_0$, the ergotropy and power can be obtained from the internal energy $\langle\hat{H}_0 \rangle_{\psi(t)}$. 

\begin{figure}[t!]
	\centering
	\includegraphics[width=\linewidth]{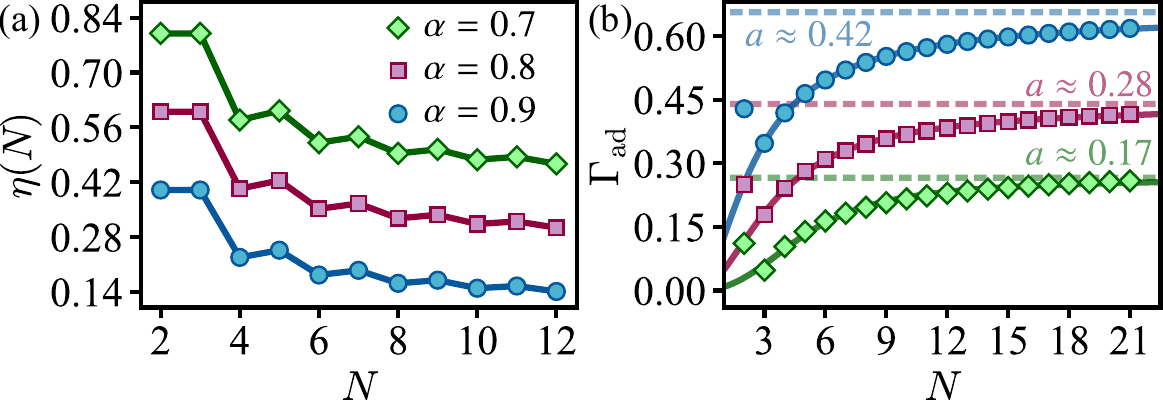}
	\caption{\textbf{Quantum advantage parameters}. (a) Driving potential and (b) power advantage of the battery charging as function of $N$, for values of $\alpha \in \{0.7,0.8,0.9\}$. We assume $g = 1.0\times 2\pi$~MHz for both graphs.}
	\label{fig:Scaling_beyond12}
\end{figure}

In order to extrapolate the behavior of the power advantage $\Gamma_\mathrm{ad}(N)$, we considered two factors: i) the behavior of $\Gamma_\mathrm{ad}(N)$ from $N=3$ to $N=21$, and ii) the results obtained in Ref.~\cite{Kim:22}, stating that the charging advantage is not extensive in the thermodynamics limit without global interactions. Based on that, we propose the scaling law
\begin{align}
	\Gamma_\mathrm{ad}(N,\{a, b, c\}) = a \arctan(b N ^{c}) , \label{Ap:Fit}
\end{align}
where the set $\{a, b, c\}$ are parameters to be adjusted. In particular, the parameter $a$ is the most relevant to our analysis because we can show that (use that $\lim_{x\rightarrow\infty}\arctan(x)=\pi/2$)
\begin{align}
	\lim_{N\rightarrow \infty} \Gamma_\mathrm{ad}(N,\{a, b, c\}) \rightarrow \frac{a \pi}{2}  ,
\end{align}
what means that $a$ gives us the asymptotic power advantage. The scaling of the quantum advantage parameters is shown in Fig.~\ref{fig:Scaling_beyond12}, with the parameter $\eta$ computed up to 13 cells, and the parameter $\Gamma_\mathrm{ad}$ computed up to $N=21$. The quantities are shown for different values of the ratio $\alpha = \Omega / g$ as function of $N$. Also, we show the curves obtained with Eq.~\eqref{Ap:Fit}, where the value of $a$ is highlighted as horizontal dashed line for each curve.

First, because driving potential ratio $\eta$ is measured up to 13 cells (see Fig.~\SubFig{fig:Scaling_beyond12}{a}), we cannot state we have genuine quantum advantage for all cases considered in Fig.~\SubFig{fig:Scaling_beyond12}{b}. However, all choices of $\alpha$ suggest we can observe genuine quantum advantage for battery sizes beyond 12 cells. In particular, the cases in which the quantum charging is much less energetically demanding (e.g. $\alpha = 0.8$ and $\alpha = 0.9$), the advantage parameter suggests a reasonable and genuine advantage of the quantum charging with respect to the classical one.

\RefB{
	\section{Entanglement generation}
	
	In this section, we discuss the generation of entanglement during the charging process. Since no entanglement is expected in the case of classical charging, our focus here is on the quantum regime. We measure an entanglement witness based on the second-order Rényi entropy~\cite{Horodecki:96PLA,Horodecki:96,Horodecki:09}.  
	
	\begin{figure*}[t!]
		\centering
		\includegraphics[width=\linewidth]{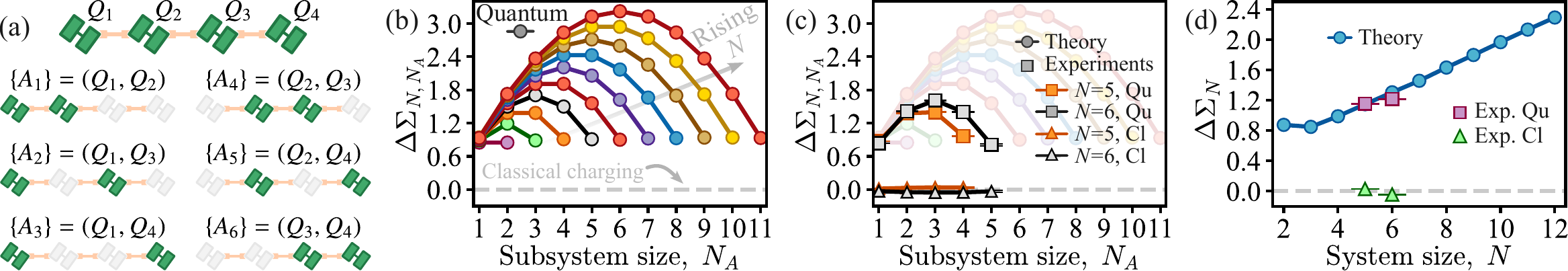}
		\caption{\RefB{\textbf{Entanglement witness via Rényi entropy}. (a) Sketch showing how we define the bi-partitions, for the example $N = 4$ and $N_{A} = 2$, to compute the bipartite, each entropy $S_{N,A_i,N_A}$ defined in Eq.~\eqref{SM:Eq:Entropy_Ai}. (b) The complete theoretical prediction of the entropy evaluated at $t \approx \Delta t_\mathrm{exp}$. (c) The experimental results for classical (triangles) and quantum charging (square) for $N=\{5,6\}$, with the theoretical curves (circles) shown in the background for comparison.  (d) Scaling of the average entropy as a function of the number of qubits in the battery, $N$. The two magenta square markers indicate the experimental data, while the blue circular markers correspond to the theoretical predictions.}}
		\label{fig:SM_Entanglement}
	\end{figure*}
	
	We consider a $N$ qubits system, partitioned into two subsystems $A$ and $B$, containing $N_A$ and $N_B$ qubits, respectively, such that $N_A + N_B = N$. The second-order Rényi entropy of subsystem $A$ is defined as
	$S_{N,A,N_A} = - \log \big[\mathrm{tr}(\rho_{A}^2)\big]$,
	where $\rho_A$ is the reduced density matrix of subsystem $A$. 
	
	Since subsystem $A$ can be chosen in multiple ways, we consider the average entropy over all possible bipartitions. Specifically, we define
	\begin{equation}
		S_{N,A_i,N_A} = - \log \big[\mathrm{tr}(\rho_{A_i}^2)\big] , \label{SM:Eq:Entropy_Ai}
	\end{equation}
	where the set $\{A_i\}$ contains all possible choices of subsystems with $N_A$ qubits. For example, as depicted in Fig.~\SubFig{fig:SM_Entanglement}{a}, for a system with $N = 4$, labeled $(Q_1, Q_2, Q_3, Q_4)$, and for bipartite systems with $N_A = 2$, we have
	$
	\{A_i\} = \left\{(Q_1,Q_2), (Q_1,Q_3), (Q_1,Q_4), (Q_2,Q_3), (Q_2,Q_4), (Q_3,Q_4)\right\}.
	$
	
	The averaged Rényi entropy is then defined as
	\begin{equation}
		S_{N,N_A} = \frac{1}{\mathrm{dim}\{A_i\}} \sum_{\{A_i\}} S_{N,A_i,N_A} ,
	\end{equation}
	where $\mathrm{dim}\{A_i\}$ is the number of distinct bipartitions of size $N_A$.  
	To study the dynamical buildup of entanglement during charging, we analyze the change in entropy,
	\begin{equation}
		\Delta \Sigma_{N,N_A} = S_{N,N_A}(\Delta t) - S_{N,N_A}(0) ,
	\end{equation}
	which quantifies the entanglement growth over a time interval $\Delta t$. We identify the presence of entanglement in the system when $\Delta \Sigma_{N,N_A} > 0$.
	
	To estimate the Rényi entropy in our setup, we employed an experimental procedure based on random measurements---see Ref.~\cite{Tiff:19} and Methods Section from Ref.~\cite{Hu:25} for further details. The maximum charging power is reached before $\Delta t =0.1 \, \mu\text{s}$, then we evaluate $\Delta \Sigma_{N,N_A}$ at $\Delta t_\mathrm{exp} \approx 0.107 \, \mu\text{s}$, for which we have Renyi entropy measurements data
	for the charging process with $N = 5$ and $N = 6$ cells.
	
	In Fig.~\SubFig{fig:SM_Entanglement}{b} we show the expected behavior of the bipartite Rényi entropy variation from $N=2$ up to $N=12$. The behavior observed in Fig.~\SubFig{fig:SM_Entanglement}{b} is experimentally verified in Fig.~\SubFig{fig:SM_Entanglement}{c} by replacing the theory data with the experimental ones when $N = 5$ and $N = 6$. Qualitatively, the experiments show the same entropy increasing behavior as the theoretical one, which indicates that entanglement is created and distributed over the system. The experimental data for the classical charging is also shown in Fig.~\SubFig{fig:SM_Entanglement}{c}. The values for entropies close to zero for all subsystems reinforce the evidence that no entanglement is created during the classical charging, as expected from the theoretical predictions. 
	
	We also investigate how the average entropy increases as a function of the system size. For this purpose, we compute the entropies for all possible partition sizes $N_A$, and define the arithmetic mean as
	\begin{align}
		\Delta \Sigma_{N} = \frac{1}{N-1} \sum_{N_A = 1}^{N-1} \Delta \Sigma_{N,N_A} ,
	\end{align}
	which represents the average entropy growth obtained by equally weighting all bipartition sizes of a system with $N$ qubits. As result, we obtain the linear growing shown in Fig.~\SubFig{fig:SM_Entanglement}{d}, where we show the theoretical data and highlight the two experimental points obtained for the cases $N = 5$ and $N = 6$ cells for the quantum charging. Also, we show the entropies in the classical charging.

	As a side note, the error estimate for the entropy is considered in our analysis. We assume noise adds a uniform ``entropy background'', proportional to the subsystem size. Therefore, from this estimate obtained for the full system entropy is then subtracted out to recover a better estimate of the true entanglement entropy according to~\cite{Hu:25}
	\begin{equation}
		S^{\mathrm{post}}_{N,A_i,N_A} = S_{N,A_i,N_A} - \frac{S_{N,A_i,N_A} N_A}{N} .
	\end{equation}
	
	This assumption is reasonably supported by previous results reported in Refs.~\cite{Elben:18}. Therefore, all theoretical data are shown without any post-correction, while the experimental data are corrected according to the above equation. In this way, we observe a good agreement of the experimental data after this systematic noise-correcting approach.
	
}

\end{document}